\documentclass[]{aa} 
\usepackage{graphicx}
\usepackage{txfonts}
\begin{document}
%
\def\hi {H\,{\sc i}}
\def\hii {H\,{\sc ii}}
\def\hdueo {H$_2$O}
\def\meth {CH$_{3}$OH}
\def\dg{$^{\circ}$}
\def\kms{km\,s$^{-1}$}
\def\ms{m\,s$^{-1}$}
\def\jyb{Jy\,beam$^{-1}$}
\def\mjyb{mJy\,beam$^{-1}$}
\def\solmass {\hbox{M$_{\odot}$}}
\def\solum {\hbox{L$_{\odot}$}} 
\def\d {$^{\circ}$}
\def\n {$n_{\rm{H_{2}}}$}
\def\kmsg{km\,s$^{-1}$\,G$^{-1}$}
\title{The properties and polarization of the \hdueo ~and \meth ~maser environment of NGC7538-IRS\,1}

\author{G.\ Surcis \thanks{Member of the International Max Planck Research School (IMPRS) for Astronomy and Astrophysics at 
the Universities of Bonn and Cologne.}  \inst{, 1}
  \and 
  W.H.T. \ Vlemmings \inst{1}
 \and
  R.M. \ Torres \inst{1}
  \and
  H.J. \ van Langevelde \inst{2,3}
  \and
  B. \ Hutawarakorn Kramer \inst{4,5}
  }

\institute{ Argelander-Institut f\"{u}r Astronomie der Universit\"{a}t Bonn, Auf dem H\"{u}gel 71, 53121 Bonn, Germany\\
 \email{gsurcis@astro.uni-bonn.de}
 \and
 Joint Institute for VLBI in Europe, Postbus 2, 79990 AA Dwingeloo, The Netherlands
 \and
 Sterrewacht Leiden, Leiden University, Postbus 9513, 2300 RA Leiden, The Netherlands
 \and
 Max-Planck Institut f\"{u}r Radioastronomie, Auf dem H\"{u}gel 69, 53121 Bonn, Germany
 \and
 National Astronomical Research Institute of Thailand, Ministry of Science and Technology, Rama VI Rd., Bangkok 10400, Thailand
  }

\date{Received ; accepted}
\abstract
{NGC7538 is a complex massive star-forming region. The region is composed of several radio continuum
sources, one of which is IRS\,1, a high-mass protostar, from which a 0.3~pc molecular bipolar outflow was detected. 
Several maser species have been detected around IRS\,1. The \meth ~masers have been suggested to trace a Keplerian-disk, while the \hdueo
~masers are almost aligned to the outflow. More recent results suggested that the region hosts a torus and potentially a disk,
but with a different inclination than the Keplerian-disk that is supposed to be traced by the \meth ~masers.}
{Tracing the magnetic field close to protostars is fundamental for determining the orientation of the disk/torus. Recent studies showed 
that during the protostellar phase of high-mass star formation the magnetic field is oriented along the outflows and around or on the
 surfaces of the disk/torus. The observations of polarized maser emissions at milliarcsecond resolution
can make a crucial contribution to understanding the orientation of the magnetic field and, consequently, the orientation
of the disk/torus in NGC7538-IRS\,1.} 
{The NRAO Very Long Baseline Array was used to measure the linear polarization and the Zeeman-splitting of the 22\,GHz \hdueo ~masers 
toward NGC7538-IRS\,1. The European VLBI Network and the MERLIN telescopes were used to measure the linear polarization and 
the Zeeman-splitting of the 6.7\,GHz \meth ~masers toward the same region.}
{We detected 17 \hdueo ~masers and 49 \meth ~masers at high angular resolution. We detected linear polarization 
emission toward two \hdueo ~masers and toward
twenty \meth ~masers. The \meth ~masers, most of which only show a core structure, seem to trace rotating and potentially infalling gas in the inner part of a 
torus. Significant Zeeman-splitting was measured in three \meth ~masers. No significant 
(3$\sigma$) magnetic field strength was measured using the \hdueo ~masers. We also propose a new 
description of the structure of the NGC7538-IRS\,1 maser region.}
{}

\titlerunning{NGC7538: magnetic field.}
\authorrunning{Surcis et al.}

\maketitle
\section{Introduction}

\indent NGC7538 is a complex massive star-forming region located in the Perseus arm of our Galaxy at a distance of 2.65 kpc 
(Moscadelli et al. \cite{mos09}). The region is composed of several clusters of infrared sources (Wynn-Williams et al. 
\cite{wyn74}) and radio continuum sources (Campbell \cite{cam84}). The brightest source is NGC7538-IRS\,1, whose central star 
has been suggested to be an O6 star of about 30~\solmass ~with systemic local standard of rest velocity $V_{\rm{lsr}}=-58$~\kms 
~(Campbell \& Thompson \cite{cam84b}; Sandell et al. \cite{san09}; Puga et al. \cite{pug10}). Several 
high-velocity molecular bipolar outflows were detected in NGC7538 and one of these is elongated 0.3~pc from IRS\,1 (position angle 
PA=140\d) 
with a velocity of 250~\kms ~and a mass of 82.8~\solmass ~(Kameya et al. \cite{kam89}; Gaume et al. \cite{gau95};
Davis et al. \cite{dav98}; Qiu et al. \cite{qiu11}). VLA continuum observations by Campbell (\cite{cam84}) indicate that the PA of the 
outflow decreases away from IRS\,1; i.e., 180\d ~at 0$\farcs$3 (0.004~pc) and 165\d ~at $2''$ (0.03~pc). Kameya et al. 
(\cite{kam89}) gave three possible interpretations of this rotation: disk precession, interaction of the flow with dense gas, 
and coupling of the gas with a large-scale magnetic field around IRS\,1. Because Sandell et al. (\cite{san09}) found that the
 collimated free-free jet (opening angle $\lesssim 30$\d) is approximately aligned with the outflow and that there is a strong accretion 
flow toward IRS\,1 (accretion rate $\sim2\times10^{-4}$~\solmass$\rm{yr^{-1}}$), IRS\,1 must be surrounded by an accretion 
disk. The morphology of the free-free emission, which is optically thick up to 100~GHz (Franco-Hern\'{a}ndez  \& Rodr\'{i}guez \cite{fra04}), 
suggests that the disk should be almost edge-on and oriented east-west (Scoville 
et al \cite{sco86}; Kameya et al \cite{kam89}; Sandell et al. \cite{san09}). A possible detection of this edge-on disk was made 
by Minier et al. (\cite{min98}), who observed a linear distribution of the brightest \meth ~masers with an inclination angle of 
about 112\d. Pestalozzi et al. (\cite{pes04}) estimated that this disk is a Keplerian disk with an outer radius of $\sim$750~AU 
and an inner radius of $\sim$290~AU by modeling the \meth ~maser emissions at 6.7 and 12.2-GHz.\\
\indent Recent results disagree with the edge-on disk traced by the \meth ~masers. The observations of the mid-infrared 
emission suggest that the radio 
continuum emission does not come from a free-free jet but that it traces the ionized gas wind from the disk surface that in the new 
scenario would be perpendicular to the CO-bipolar outflow with a disk inclination angle $i=32$\d~ (De~Buizer \& Minier \cite{bui05}). 
Klaassen et al. (\cite{kla09}) observed two warm gas tracers (SO$_{2}$ and OCS) toward NGC7538-IRS\,1 with the Submillimeter 
Array (SMA). Although the region was unresolved, they found a velocity gradient consistent with the \meth ~maser velocities 
and perpendicular to the large-scale molecular bipolar outflow. This rotating gas might indicate that there is a torus (with 
an angular size of about $2$ arcsec; i.e., $\sim$5300~AU at 2.65~kpc) surrounding the smaller accretion disk proposed by De Buizer \& Minier
(\cite{bui05}).\\
\indent In addition, Krauss et al. (\cite{kra06}) proposed two possible scenarios for NGC7538-IRS\,1 by considering both the linear distribution 
of the 6.7-GHz \meth ~masers detected previously (e.g., Pestalozzi et al. \cite{pes04}, De Buizer \& Minier \cite{bui05})
 and the asymmetry in the near infrared (NIR) images observed by them. In their most likely scenario, called 
``Scenario B'', the \meth ~masers trace the edge-on disk as suggested by Pestalozzi et al. (\cite{pes04}) and the asymmetry
might be caused by the precession of the jet. In the ``Scenario A'' the \meth ~masers do not trace a disk but an outflow cavity as proposed by
De Buizer \& Minier (\cite{bui05}), where the detected asymmetry might simply reflect the innermost walls of this cavity.\\
\begin{figure*}[th!]
\centering
\includegraphics[width = 13 cm]{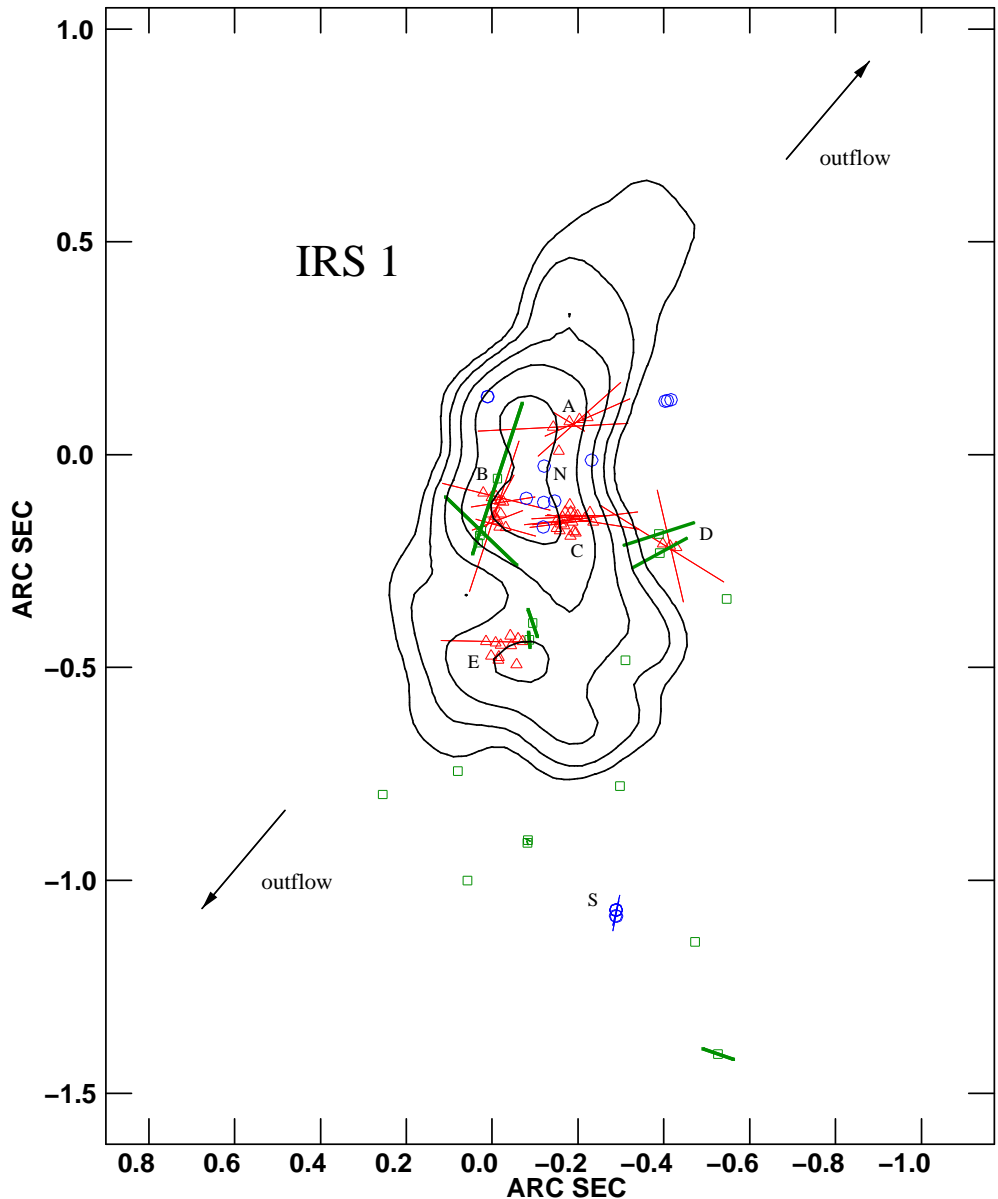}
\caption{Positions of water, methanol, and hydroxyl maser features superimposed on the 2\,cm continuum contour map of NGC7538-IRS\,1 
observed with the 
VLA in 1986 (Franco-Hern\'{a}ndez  \& Rodr\'{i}guez \cite{fra04}). Contours are 1, 2, 4, 8, 16 $\times$1.3 \mjyb. The (0,0) 
position is $\alpha_{2000}=23^{\rm{h}}13^{\rm{m}}45^{\rm{s}}\!.382$ and $\delta_{2000}=+61^{\circ}28'10''\!\!.441$. Blue circles
 and red triangles indicate the positions of the \hdueo ~and \meth ~maser features, respectively, with their linear polarization 
vectors (60 mas correspond to a linear polarization fraction of 1\%). Green boxes indicate the positions of the OH masers detected by
Hutawarakorn \& Cohen (\cite{hut03}) with their linear polarization vectors (thick lines, 6~mas correspond to 1\%). The two arrows indicate 
the direction of the bipolar outflow ($\sim$140\d).}
\label{pos}
\end{figure*}
\indent Besides the \meth ~masers, other maser species were detected around NGC7538-IRS\,1: OH, \hdueo, NH$_{3}$, and H$_{2}$CO 
(e.g., Hutawarakorn \& Cohen \cite{hut03}; Galv\'{a}n-Madrid et al. \cite{gal10}; Gaume et al. \cite{gau91}; Hoffman et al.
\cite{hof03}). The OH masers are located southward and show no obvious disk structure or relation to the outflow direction 
(Hutawarakorn \& Cohen \cite{hut03}). H$_{2}$CO and \hdueo ~masers are located near the center of the continuum emission, and the \hdueo 
~masers are also almost aligned with the outflow (Galv\'{a}n-Madrid et al. \cite{gal10}). The 6.7 and 12.2-GHz \meth ~masers show a 
cone shape that opens to the north-west (Minier et al. \cite{min00}). \\
\indent So far, the magnetic field structure in NGC7538-IRS\,1 has been studied using submillimeter imaging polarimetry (Momose 
et al. \cite{mom01}) and OH maser emission (Hutawarakorn \& Cohen \cite{hut03}). Although the polarization vectors are locally 
disturbed, at an angular resolution of 14~arcsec the magnetic field directions agree with the direction of the outflow 
(Momose et al. \cite{mom01}), while at milliarcsecond (mas) resolution the OH maser observations indicate a magnetic field oriented 
orthogonal to the outflow (Hutawarakorn \& Cohen \cite{hut03}). Because \hdueo ~and \meth ~masers were detected close to the center of 
the continuum emission, i.e. to the protostar, it is worthwhile investigating their linear and circular polarization emissions. 
As shown by Surcis et al. (\cite{sur09}, \cite{sur11}) for the massive star-forming region W75N, the polarization observations 
of the two maser species can depict a reasonable scenario for the magnetic field close to massive protostars. Moreover, the direction 
and strength of the magnetic field might help to decide the debate about the orientation of the disk.\\
\indent Here we present Very Long Baseline Array (VLBA) observations of \hdueo ~masers, Multi-Element Radio Linked 
Interferometer network (MERLIN) and European VLBI Network (EVN) observations of \meth~ masers in full polarization toward 
NGC7538-IRS\,1. In Sect.~\ref{res} we show our results obtained by studying the linear and circular 
polarization of \hdueo ~and \meth ~maser emissions in a similar way as Vlemmings et al. (\cite{vle06a}, \cite{vle10}) 
and Surcis et al. (\cite{sur09}, \cite{sur11}) for Cepheus~A and W75N, respectively. In Sect.~\ref{dis} we discuss our results and  
attempt to disentangle the complex morphology in NGC7538-IRS\,1. 
\begin {table*}[th!]
\caption []{All 22 GHz \hdueo ~maser features detected in NGC7538-IRS\,1.} 
\begin{center}
\scriptsize
\begin{tabular}{ l c c c c c c c c c c c c c}
\hline
\hline
\,\,\,\,\,(1)&(2)& (3)     & (4)       & (5)           & (6)            & (7)                 & (8)         & (9)        & (10)                    & (11)                    & (12)            & (13)   \\
Maser    & Group & RA      & Dec       & Peak flux     & $V_{\rm{lsr}}$ & $\Delta v\rm{_{L}}$ &$P_{\rm{l}}$ &  $\chi$    & $\Delta V_{\rm{i}}^{a}$ & $T_{\rm{b}}\Delta\Omega^{a}$& $P_{\rm{V}}^{b}$& $\theta^{c}$\\
         &       &offset   & offset    & Density(I)    &                &                     &             &	         &                         &                         &                 &              \\ 
	 &       &(mas)    & (mas)     & (\jyb)        &  (\kms)        &      (\kms)         & (\%)        &   (\d)     & (\kms)                  & (log K sr)              & (\%)    & (\d)       \\
\hline
W01	 & N     &-185.2131& 141.908   & $17.11\pm0.07$& -57.8          & 	0.41	       &$-$          & $-$        & $-$                     & $-$                     &	$-$	   & $-$ \\
W02	 & N     &-176.4676& 139.422   & $1.00\pm0.02$ & -58.3          & 	0.32	       &$-$          & $-$        & $-$ 		    & $-$	              &	$-$	   & $-$ \\
W03	 & N     &-170.9744& 139.088   & $0.48\pm0.02$ & -58.3          & 	0.27	       &$-$          & $-$        & $-$     		    & $-$	     	      &	$-$	   & $-$         \\
W04	 & S     &-57.6929 & -1071.104 & $0.90\pm0.01$ & -72.2          & 	0.59	       &$-$          & $-$        & $-$                     & $-$                     &	$-$	   & $-$ \\
W05	 & S     &-57.4470 & -1056.554 & $0.48\pm0.01$ & -57.3          &	0.47	       &$-$          & $-$        & $-$                     & $-$                     &	$-$	   & $-$ \\
W06	 & S     &-57.3376 & -1056.630 & $36.43\pm0.02$& -70.3          &	1.15	       &$1.2\pm0.1$  & $-13.0\pm1.3$& $-$                   & $-$                     &	$-$	   & $-$ \\
W07	 & S     &-57.1463 & -1057.065 & $16.99\pm0.01$& -73.3          & 	0.77	       &$1.2\pm0.1$  & $-10.8\pm0.5$& $3.4^{+0.1}_{-0.3}$   & $9.1^{+0.3}_{-0.1}$     &	$0.37\pm0.13$& $81^{+9}_{-13}$\\
W08	 & S     &-56.7637 & -1056.586 & $0.15\pm0.01$ & -67.3          & 	1.46	       &$-$          & $-$        & $-$                     & $-$     		      &	$-$	   & $-$ \\
W09	 & S     &-56.4631 & -1070.080 & $0.48\pm0.01$ & -71.6          & 	0.64	       &$-$  	     & $-$	  & $-$		            & $-$     		      &	$-$	   & $-$ \\
W10	 & N     &0        & 0         & $7.34\pm0.03$ & -60.6          & 	0.56	       &$-$          & $-$        & $-$		            & $-$ 		      &	$-$	   & $-$ \\
W11	 & N     &85.5692 & -96.002    & $0.36\pm0.02$ & -59.7          & 	0.44	       &$-$  	     & $-$	  & $-$		            & $-$ 		      &	$-$	   & $-$ \\
W12	 & N     &110.0019 & -13.538   & $0.20\pm0.03$ & -60.2          & 	0.83	       &$-$  	     & $-$ 	  & $-$                     & $-$  		      &	$-$	   & $-$ \\
W13	 & N     &111.6690 & -99.499   & $9.31\pm0.03$ & -60.1 	        & 	0.47	       &$-$  	     & $-$        & $-$		            & $-$  		      &	$-$	   & $-$ \\
W14	 & N     &112.8442 & -156.492  & $0.25\pm0.03$ & -60.1          &	0.42	       &$-$  	     & $-$	  & $-$   		    & $-$     		      &	$-$	   & $-$ \\
W15	 & N     &151.9803 & -90.046   & $1.04\pm0.03$ & -59.9          &	0.40	       &$-$ 	     & $-$	  & $-$ 		    & $-$   		      &	$-$	   & $-$ \\
W16	 & N     &242.0042 & 148.947   & $0.13\pm0.01$ & -43.3		& 	0.62	       &$-$  	     & $-$	  & $-$		            & $-$		      &	$-$	   & $-$ \\
W17	 & N     &242.7694 & 149.715   & $0.14\pm0.01$ & -43.8       	& 	0.59	       &$-$  	     & $-$	  & $-$		            & $-$    		      &	$-$	   & $-$ \\
\hline
\end{tabular}
\end{center}
\scriptsize{\textbf{Notes.} $^{(a)}$ The best-fitting results obtained by using a model based on the radiative transfer theory of \hdueo 
~masers for $\Gamma+\Gamma_{\nu}=1$s$^{-1}$ (Surcis et al. \cite{sur11}). The errors were determined by analyzing the full probability 
distribution function.
$^{(b)}$ The percentage of circular polarization is given by $P_{\rm{V}}=(V_{\rm{max}}-V_{\rm{min}})/I_{\rm{max}}$.
$^{(c)}$The angle between the magnetic field and the maser propagation direction is determined by using the observed $P_{\rm{l}}$ 
and the fitted emerging brightness temperature. The errors were determined by analyzing the full probability distribution function.}
\label{poltw}
\end{table*}
\section{Observations and analysis}\label{obssect}
\subsection{22 GHz VLBA data}
We observed the star-forming region NGC7538-IRS1 in the 6$\rm{_{16}}$-5$\rm{_{23}}$ transition of \hdueo ~(rest frequency: 
22.23508 GHz) with the NRAO\footnote{The National Radio Astronomy Observatory (NRAO) is a facility of the National Science 
Foundation operated under cooperative agreement by Associated Universities, Inc.} VLBA on November 21, 2005. The 
observations were made in full polarization spectral mode using four overlapped baseband filters of 1 MHz to cover a 
total velocity range of $\approx50$ \kms. Two correlations were performed. One with 128 channels to generate all four 
polarization combinations (RR, LL, RL, LR) with a channel width of 7.8 kHz (0.1 \kms). The other one with high spectral 
resolution (512 channels; 1.96 kHz=0.027 \kms), which only contains the circular polarization combinations (LL, RR), to be able 
to detect Zeeman-splitting of the \hdueo ~maser across the entire velocity range. Including the overheads, the total observation 
time was 8~h.\\
\indent The data were edited and calibrated using the Astronomical Image Processing System (AIPS) following the method of Kemball 
et al. (\cite{kem95}). The bandpass, the delay, the phase, and the polarization calibration were performed on the calibrator 
J0359+5057. The fringe-fitting and the self-calibration were performed on the brightest maser feature (W06 in Table \ref{poltw}). 
All calibration 
steps were initially performed on the dataset with modest spectral resolution after which the solutions, with the exception of 
the bandpass solutions that were obtained separately, were copied and applied to the high spectral resolution dataset. Stokes 
\textit{I}, \textit{Q}, and \textit{U} data cubes (4 arcsec~$\times$~4 arcsec, rms~$\approx7$~\mjyb) were created using the AIPS task IMAGR 
(beam-size 1.0~mas~$\times$~0.4 mas) from the modest spectral resolution dataset, while the \textit{I} and \textit{V} 
 cubes (rms~$\approx8$~\mjyb) where imaged from the high spectral resolution dataset and for the same fields. The \textit{Q} and 
\textit{U} cubes were combined to produce cubes of polarized intensity and polarization angle. Because these observations were 
obtained between two VLA polarization calibration observations\footnote{http://www.aoc.nrao.edu/~smyers/calibration/}
  made by the NRAO in 2009, during which 
the linear polarization angle of J0359+5057 was constant at -86\dg$\!\!$.7, we were able to estimate the polarization angles with 
a systemic error of no more than $\sim$3\dg.\\
\indent We identified the \hdueo ~maser features using the process described in Surcis et al. (\cite{sur11}). 
Here the program called ``maser finder'' is used. This program is able to search maser spots, velocity channel by velocity channel. We
identified a maser feature when three or more maser spots coincide spatially (within a box 2 by 2 pixels) and each of them 
appeared in consecutive velocity channels. In Table~\ref{poltw} only the brightest spot of each series of maser spots that fulfill the
criteria described above are reported.
The two maser features 
that show linear polarization emission were fitted using a full radiative transfer method code based on the
models for \hdueo ~masers of Nedoluha \& Watson (\cite{ned92}). They solved the transfer equations for the polarized radiation of 22~GHz
\hdueo ~masers in the presence of a magnetic field which causes a Zeeman-splitting that is much smaller than spectral line breadth. 
The fit provides the emerging brightness temperature ($T_{\rm{b}}\Delta\Omega$) and the intrinsic thermal linewidth ($\Delta V_{\rm{i}}$). 
See Vlemmings et al. (\cite{vle06a}) and Surcis et al. (\cite{sur11}) for more details. We modeled the observed linear polarized and 
total intensity maser spectra by gridding $\Delta V_{\rm{i}}$ from 0.5 to 3.5~\kms, in steps of 0.025~\kms, by using a least-squares 
fitting routine. From the fit results we were able to determine the best values of the angle between the maser propagation 
direction and the magnetic field ($\theta$). By considering the values of $T_{\rm{b}}\Delta\Omega$, $\Delta V_{\rm{i}}$, and $\theta$
 we were also able to estimate the saturation state of the \hdueo ~maser features. 
\begin{figure*}[th!]
\centering
\includegraphics[width = 9 cm]{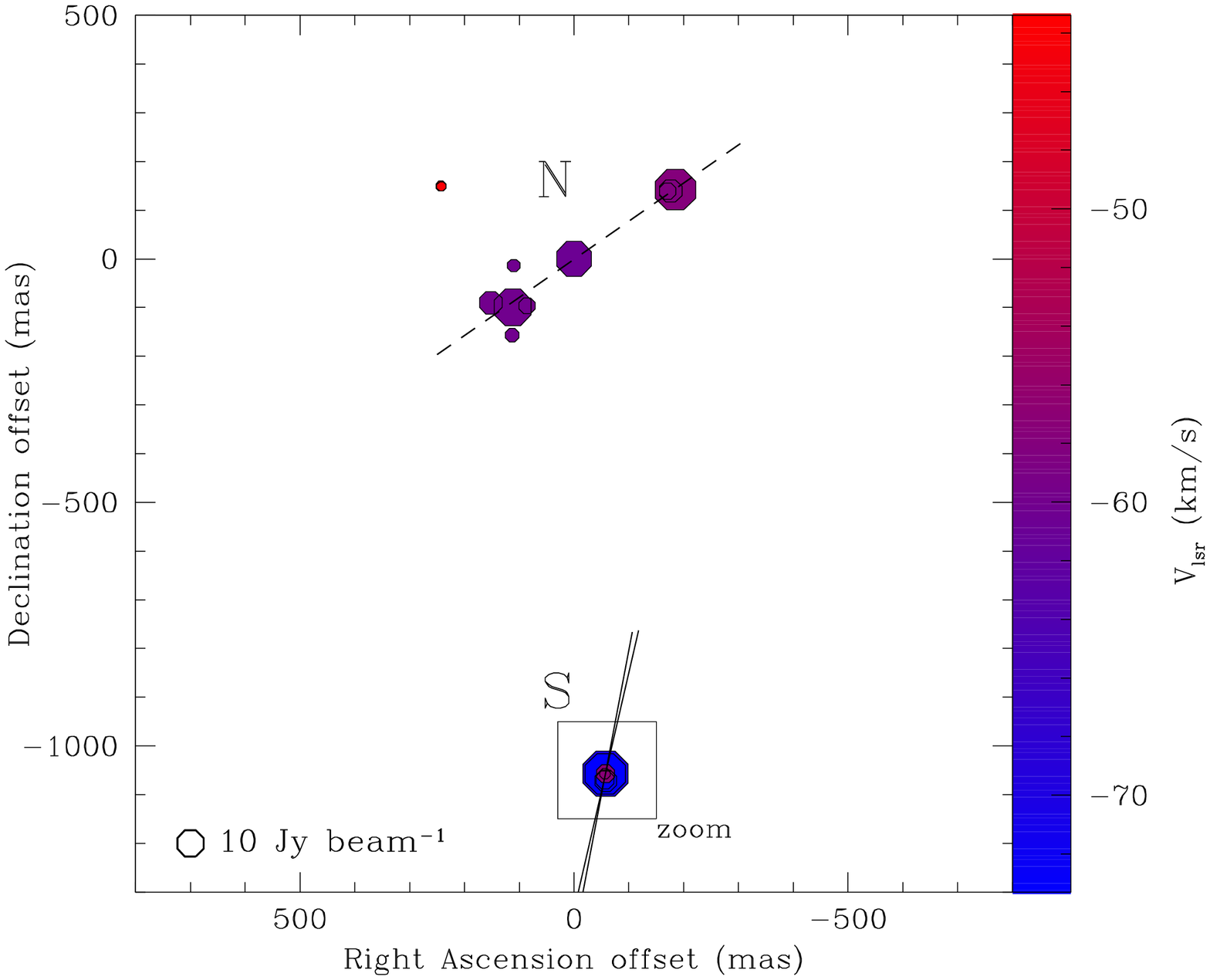}
\includegraphics[width = 9 cm]{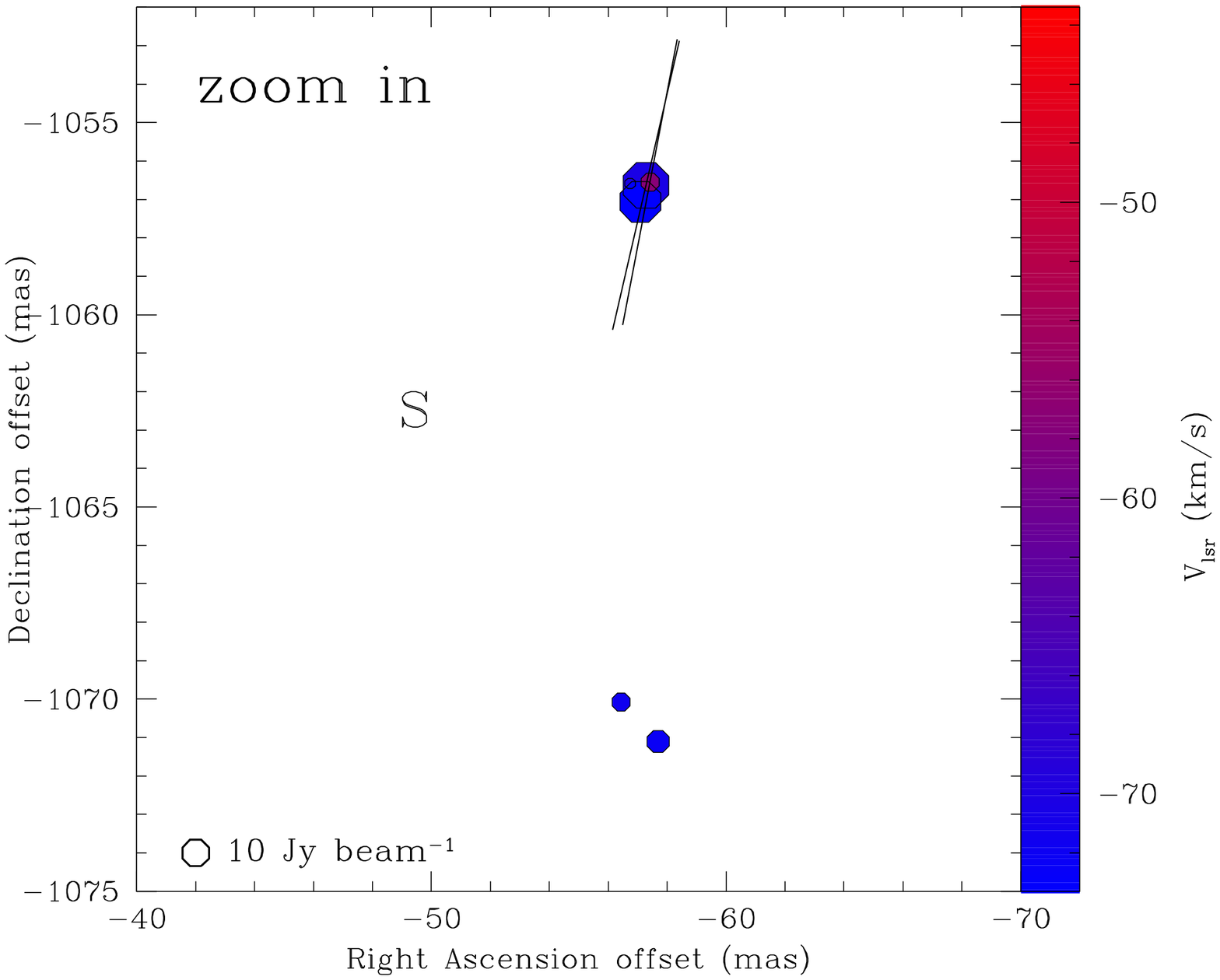}
\caption{Left panel: a close-up view of the \hdueo ~maser features around NGC7538-IRS1. Right panel: a zoom-in view of the boxed 
region of the left panel. The octagonal symbols are the identified maser features in the present work scaled logarithmically according 
to their peak flux density. The maser LSR radial velocity is indicated by color. A 10~\jyb ~symbol is plotted for illustration in 
both panels. 
The reference position is $\alpha_{2000}=23^{\rm{h}}13^{\rm{m}}45^{\rm{s}}\!.350$ and
 $\delta_{2000}=+61^{\circ}28'10''\!\!.428$, which was estimated as described in Sect.~\ref{res}. 
The linear polarization vectors, scaled logarithmically according to polarization fraction $P_{\rm{1}}$ 
(in Table~\ref{poltw}), are overplotted. The dashed line (PA=-52\d) is the result of the best linear fit of the \hdueo ~masers of 
group N (features W16 and W17 were not included).The synthesized beam is 1.0~mas~$\times$~0.4~mas. }
\label{wat}
\end{figure*}
\subsection{6.7 GHz MERLIN data}
To detect the polarization of the \meth ~maser emission at 6668.518\,MHz ($5_{1}-6_{0}\,\rm{A^{+}}$) we observed 
NGC7538-IRS1 with six of the MERLIN\footnote{MERLIN is operated by the University of Manchester as a National Facility of the 
Science and Technology Facilities Council} telescopes in full polarization spectral mode on December 28, 2005. The observation
 time was $\sim2$~h, including overheads on the calibrators 2300+638, 3C84 and 3C286. We used a 250\,kHz bandwidth ($\sim11$\kms) 
with 256 channels (velocity resolution $\sim0.04$~\kms) centered on the source velocity $V_{\rm{lsr}}=-56.1$~\kms. For 
calibration purposes, the continuum calibrators were observed with the 16\,MHz wide-band mode. Both 3C84 and 3C286 were also 
observed in the narrow-band spectral line configuration and were used to determine the flux and bandpass calibration solutions. 
The data were edited and calibrated using AIPS. The calibrator 3C84 was used to determine the phase offset between the wide- and 
narrow-band set-up. 
Instrumental feed polarization was determined using the unpolarized calibrator 3C84, and the polarization angle was calibrated 
using 3C286. Using one of the strongest isolated maser features, we were able to self-calibrate the data in right- and left-circular 
polarization separately. After calibration, the antenna contributions were reweighed according to their sensitivity at 5\,GHz and 
their individual efficiency. Stokes \textit{I}, \textit{Q}, \textit{U} data cubes (5.12~arcsec $\times$ 5.12~arcsec, rms~$\approx10$~\mjyb) 
were created (beam-size 47\,mas $\times$ 34\,mas).\\
\indent The \meth ~maser features were also identified through the identification process mentioned above (Surcis et al. \cite{sur11}).
If the program finds a group of maser spots that are not exactly spatially coincident but show a continuum linear distribution with a
clear velocity gradient, we report in the corresponding table only the brightest maser spot of the group (e.g. maser feature M06).
We were unable to identify weak \meth ~maser features ($<1$~\jyb) close to the brightest ones because of the dynamic range limits. 
To determine $T_{\rm{b}}\Delta\Omega$, $\Delta V_{\rm{i}}$, and $\theta$, we adapted the code used 
for 22\,GHz \hdueo ~masers to model the 6.7\,GHz \meth \,masers, which has successfully been used by Vlemmings et al. 
(\cite{vle10}) for the \meth ~masers in Cepheus\,A. We modeled the observed linear polarized and total intensity \meth ~maser feature
spectra by gridding $\Delta V_{\rm{i}}$ from 0.5 to 2.6~\kms, in steps of 0.05~\kms, by using a least-squares fitting routine.

\subsection{6.7 GHz EVN data}
NGC7538-IRS1 was also observed at 6.7\,GHz (\meth) in full polarization spectral mode with nine of the EVN\footnote{The European VLBI Network
is a joint facility of European, Chinese, South African and other radio astronomy institutes funded by their national research councils.}
 antennas (Jodrell2, Cambridge, Efelsberg, Onsala, Medicina, Torun, Noto, Westerbork, and the new joint antenna Yebes-40\,m), for a total 
observation time of 5\,h on November 3, 2009 (program code ES063B). The bandwidth was 2\,MHz, providing 
a velocity range of $\sim100$~\kms. The data were correlated using 1024 channels to generate all four polarization 
combinations (RR, LL, RL, LR) with a spectral resolution of 1.9\,kHz ($\sim$0.1~\kms).\\
\indent The data were edited and calibrated using AIPS. The bandpass, the delay, the phase, and the polarization calibration were 
performed on the calibrator 3C286. Fringe-fitting and self-calibration were performed on the brightest maser feature (E26 in 
Table~\ref{poltm}). Then we imaged the \textit{I}, \textit{Q}, \textit{U}, \textit{RR}, and \textit{LL} cubes (2~arcsec~$\times$~2~arcsec, 
rms~$\approx8$~\mjyb) using the AIPS task IMAGR (beam-size 6.3~mas $\times$ 4.9~mas). The \textit{Q} and \textit{U} cubes were combined 
to produce cubes of polarized intensity and polarization angle. \\
\indent In this case, because the dynamic range of our EVN observations was better than that of MERLIN observations, we were able to 
identify \meth ~maser features with a peak flux density of less than 1~\jyb ~with the ``maser finder'' program. The maser emission was fitted by 
using the adapted code for \meth ~masers, but with a grid of $\Delta V_{\rm{i}}$ ranging from 0.5 to 1.95~\kms.
 We were able to determine the Zeeman-splitting from the 
cross-correlation between the RR and LL spectra, which was successfully used by Surcis et al. (\cite{sur09}) for the polarized \meth 
~maser emission detected in W75N. The dynamic range of the RR and LL cubes decreases close to the strongest maser emission of each 
group because of the residual calibration errors. As a result, we were not able to determine the 
Zeeman-splitting ($\Delta V_{\rm{Z}}$) for the features with a peak flux density of less than 1.8~\jyb.
\begin {table*}[th!]
\caption []{All 6.7 GHz \meth ~maser features detected in NGC7538-IRS\,1 with MERLIN.} 
\begin{center}
\scriptsize
\begin{tabular}{ l c c c c c c c c c c c c c }
\hline
\hline
\,\,\,\,\,(1)&(2)& (3)      & (4)      & (5)        & (6)            & (7)                 & (8)          & (9)          & (10)                    & (11)                        & (12)\\
Maser    & group & RA$^{a}$& Dec$^{a}$& Peak flux     & $V_{\rm{lsr}}$ & $\Delta v\rm{_{L}}$ &$P_{\rm{l}}$ &  $\chi$    & $\Delta V_{\rm{i}}^{b}$ & $T_{\rm{b}}\Delta\Omega^{b}$ &  $\theta^{c}$\\
         &       & offset & offset & Density(I)    &                &                     &             &	          &                         &                            &             \\ 
	 &       & (mas)  & (mas)  & (\jyb)        &  (\kms)        &      (\kms)         & (\%)        &   (\d)     & (\kms)                  & (log K sr)                      &  (\d)       \\
\hline
M01	 &   D   & -256   & -275   & $22.91\pm0.01$& -60.68         & 0.86	          &$7.1\pm1.4$  & $+39\pm8$  & $ <0.55^d$              & $ >12^d$                        & $-$ \\
M02	 &   A   & -115   & 47     & $1.70\pm0.02$ & -55.94         & 0.45	          &$-$          & $-$        & $-$                     & $-$                             & $-$ \\
M03	 &   A   & -74    & 22     & $23.53\pm0.01$& -57.30         & 0.36	          &$1.3\pm0.6$  & $-47\pm3$  & $1.20^{+0.08}_{-0.15}$  & $8.7^{+0.1}_{-1.3}$             & $90^{+53}_{-53}$\\
M04	 &   C   & -73    & -215   & $51.32\pm0.01$& -61.38         & 0.37	          &$1.3\pm0.4$  & $-42\pm4$  & $-$                     & $-$                             & $-$ \\
M05	 &   C   & -28    & -222   & $50.53\pm0.01$& -60.81         & 0.85	          &$0.7\pm0.4$  & $+41\pm8$  & $ 2.20^{+0.06}_{-0.13}$ & $8.4^{+0.3}_{-1.0}$             & $76^{+11}_{-46}$ \\
M06	 &   A   & 0      & 0      & $172.00\pm0.02$&-56.07         & 0.63	          &$1.0\pm0.4$  & $+45\pm60$ & $ <0.95^d$              & $ >12^d$	                 & $-$ \\
M07	 &   B   & 59     & -287   & $3.88\pm0.01$ & -58.44         & 0.52	          &$1.5\pm0.9$  & $-29\pm7$  & $-$                     & $-$                             & $-$ \\
M08	 &   B   & 98     & -186   & $1.11\pm0.01$ & -58.70         & 0.54	          &$-$          & $-$        & $-$                     & $-$                             & $-$ \\
M09	 &   E   & 87     & -559   & $3.54\pm0.01$ & -59.05         & 0.31	          &$1.6\pm0.2$  & $+33\pm6$  & $1.40^{+0.08}_{-0.12}$  & $8.8^{+0.4}_{-0.3}$             & $87^{+4}_{-20}$ \\
M10	 &   E   & 125    & -546   & $1.08\pm0.01$ & -58.83         & 0.42	          &$-$          & $-$        & $-$                     & $-$                             & $-$ \\
M11	 &   B   & 127    & -207   & $24.41\pm0.01$& -57.61         & 0.70	          &$0.5\pm0.1$  & $-23\pm15$ & $ < 0.90^d$             & $ >12^d$                        & $-$ \\
M12	 &   E   & 154    & -503   & $61.83\pm0.01$& -58.09         & 0.30	          &$0.4\pm0.1$  & $+10\pm5$  & $1.65^{+0.05}_{-0.08}$  & $8.2^{+0.7}_{-0.3}$             & $72^{+10}_{-44}$ \\
M13	 &   B   & 159    & -157   & $9.71\pm0.02$ & -56.33         & 0.40	          &$1.6\pm1.3$  & $+18\pm13$ & $-$                     & $-$                             & $-$ \\
\hline
\end{tabular}
\end{center}
\scriptsize{
\textbf{Notes.} $^{(a)}$ The absolute reference position is $\alpha_{2000}=23^{\rm{h}}13^{\rm{m}}45^{\rm{s}}\!.362$ and
 $\delta_{2000}=+61^{\circ}28'10''\!\!.506$. 
$^{(b)}$ The best-fitting results obtained by using a model based on the radiative transfer theory of \meth ~masers for 
$\Gamma+\Gamma_{\nu}=1$s$^{-1}$ (Vlemmings et al. \cite{vle10}). The errors were determined by analyzing the full probability distribution 
function. 
$^{(c)}$The angle between the magnetic field and the maser propagation direction is determined by using the observed $P_{\rm{l}}$ 
and the fitted emerging brightness temperature. The errors were determined by analyzing the full probability distribution function. 
$^{(d)}$Because of the low angular resolution the code is able to give an upper limit for $\Delta V_{\rm{i}}$ and a lower limit
for $T_{\rm{b}}\Delta\Omega$.}
\label{poltmer}
\end{table*}
\section{Results}
\label{res}
In Fig.~\ref{pos} we show the \hdueo ~(blue circles) and \meth ~maser features (red triangles) detected by us and superimposed 
on the 2~cm continuum contour map of NGC7538-IRS1 observed with the VLA in 1986 (Franco-Hern\'{a}ndez \& Rodr\'{i}guez 2004). 
Because the continuum observations were made 20.3~yr before our observations, we have shifted the continuum map by -50 mas in both 
directions according to the proper motion, $\mu_{\alpha}=-2.45$~mas~yr$^{-1}$ and $\mu_{\delta}=-2.45$~mas~yr$^{-1}$, measured by 
Moscadelli et al. (\cite{mos09}). 
The direction of the large-scale molecular bipolar outflow is also shown. Because we did not have absolute positions of the \hdueo ~maser 
features,
we estimated the offset of one common \hdueo ~maser feature detected both in the VLA observations of Galv\`{a}n-Madrid 
et al. (\cite{gal10}) and by us (features M3 and W01, respectively). All positions of the \hdueo ~maser features were shifted
according to this offset.
We did indeed have absolute positions of the \meth ~maser features at low-angular resolution, but not at high-angular resolution. 
Hence, to overlay these maser features to the continuum, we matched the brightest \meth ~maser feature detected with the EVN and 
the brightest one detected with MERLIN (M06, absolute position $\alpha_{2000}=23^{\rm{h}}13^{\rm{m}}45^{\rm{s}}\!.362$ and
 $\delta_{2000}=+61^{\circ}28'10''\!\!.506$). All \meth ~maser features were shifted accordingly. Thus, based on the resolution of the VLA
\hdueo ~and MERLIN \meth ~maser observations the uncertainties of \hdueo ~and \meth ~maser features absolute positions are 50~mas 
(Galv\`{a}n-Madrid et al. \cite{gal10}) and 10 ~mas, respectively.
\begin{figure}[h!]
\centering
\includegraphics[width = 8.8 cm]{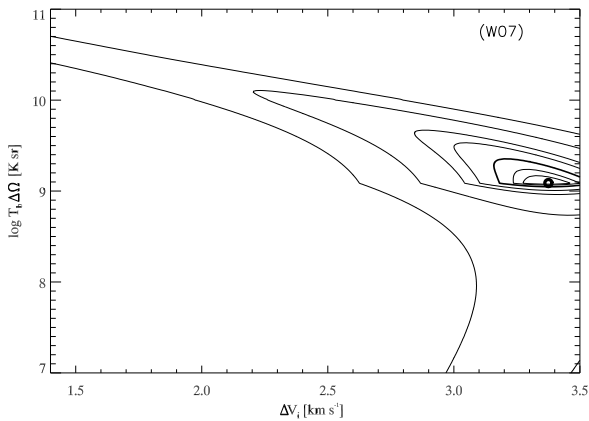}
\caption{Results of the full radiative transfer $\chi^{2}$-model fits for the \hdueo ~maser feature W07. The fit yields the 
emerging maser brightness temperature $T_{\rm{b}}\Delta\Omega$ and the intrinsic maser thermal linewidth $\Delta V_{\rm{i}}$. Contours 
indicate the significance intervals $\Delta \chi^{2}$=0.25, 0.5, 1, 2, 3, 7, with the thick solid contours indicating 1$\sigma$ 
and 3$\sigma$ areas (see Vlemmings et al. \cite{vle06a} and Surcis et al. \cite{sur11} for more details).}
\label{watf}
\end{figure}
\subsection{\hdueo~ masers}
We detected 17 22-GHz \hdueo ~maser features with the VLBA associated to NGC7538-IRS1 (named W01-W17 in Table~\ref{poltw}). No 
\hdueo ~maser emission with a peak flux density ($I$) less than 0.13 \jyb ~is detected even though our channel rms is significantly less.\\
\indent The \hdueo ~maser features can be divided into two groups, N and S, that are composed of 11 and 6 \hdueo ~maser 
features, respectively. 
Group N is located at the center of the continuum emission and it shows a linear distribution with a position angle of 
--52\d ~(see Fig.~\ref{wat}). The local standard-of-rest velocities ($V_{\rm{lsr}}$) of group N are between $-60.2$~\kms ~and $-43.3$~\kms. 
Excluding W16 and W17, the range is $-60.2$~\kms$<V_{\rm{lsr}}^{\rm{N}}<-58.3$~\kms. Group S is located about 1$''$ southward from 
group N and its velocity range is $-73.3$~\kms$<V_{\rm{lsr}}^{\rm{S}}<-57.3$~\kms, which is more blue-shifted.\\
\indent Linear polarization is detected in 2 \hdueo ~maser features that belong to group S (W06 and W07). Their linear 
polarization fraction (column 7 of Table~\ref{poltw}) is  $P_{\rm{l}}\approx1$\% and the weighted linear polarization angles is 
$\langle\chi_{\rm{H_2O}}^{\rm{S}}\rangle\approx -11$\d. The full radiative transfer method code for \hdueo ~masers was
 able to fit the feature W07 and the results are given in column 10 and 11 of Table~\ref{poltw} and in  
Fig.~\ref{watf}. The emerging brightness temperature  and the intrinsic thermal linewidth are 
$10^9$~K~sr and 3.4~\kms, respectively. By considering $T_{\rm{b}}\Delta\Omega$ and the observed $P_{\rm{l}}$ we determined 
$\theta_{\rm{H_2O}}^{\rm{B}}=81$\d, indicating that the maser is operating in a regime where the magnetic field is close to perpendicular
to the propagation of the maser radiation. No significant circular polarization emission is detected. 
\begin{figure*}[th!]
\centering
\includegraphics[width = 9 cm]{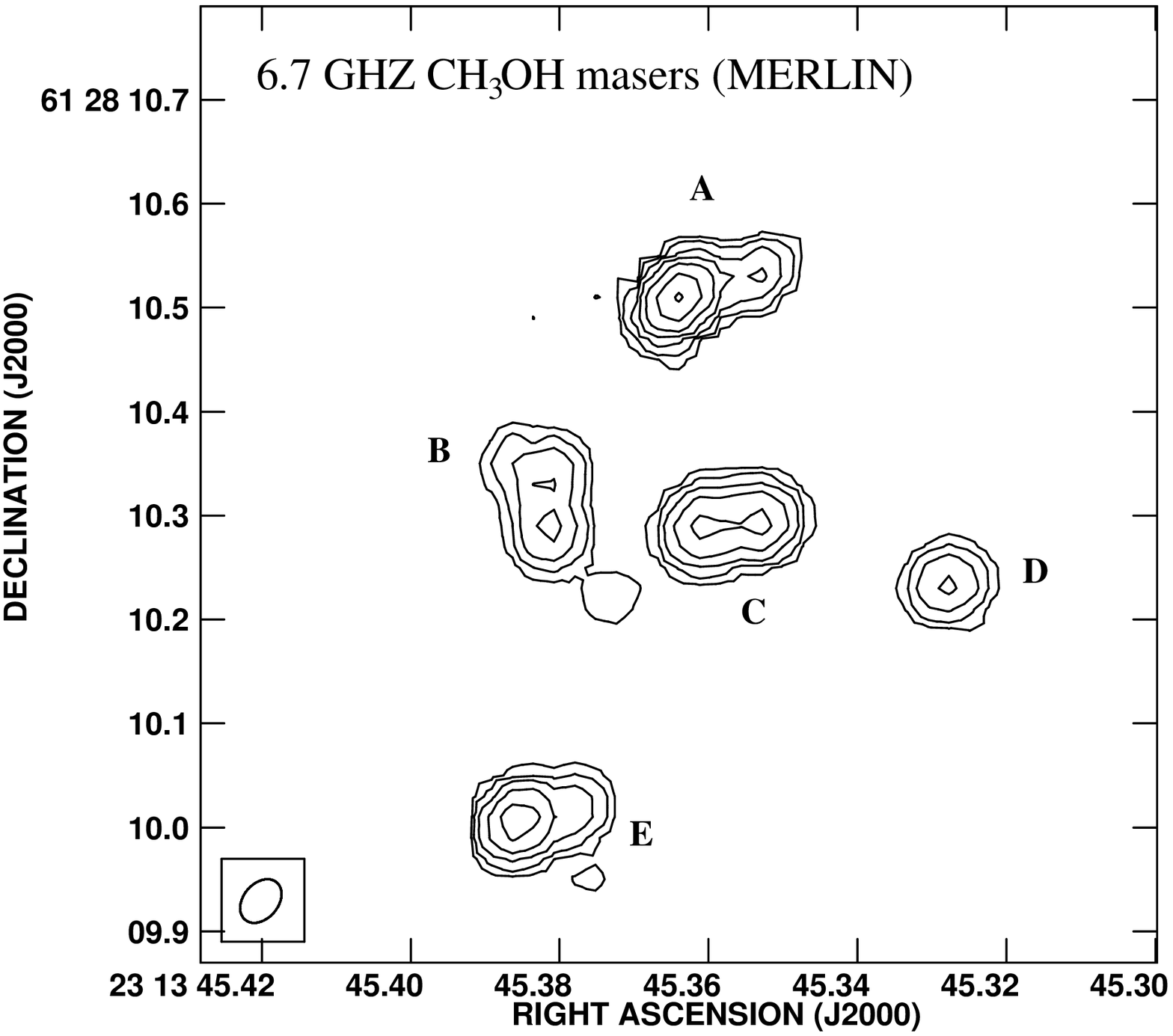}
\includegraphics[width = 9 cm]{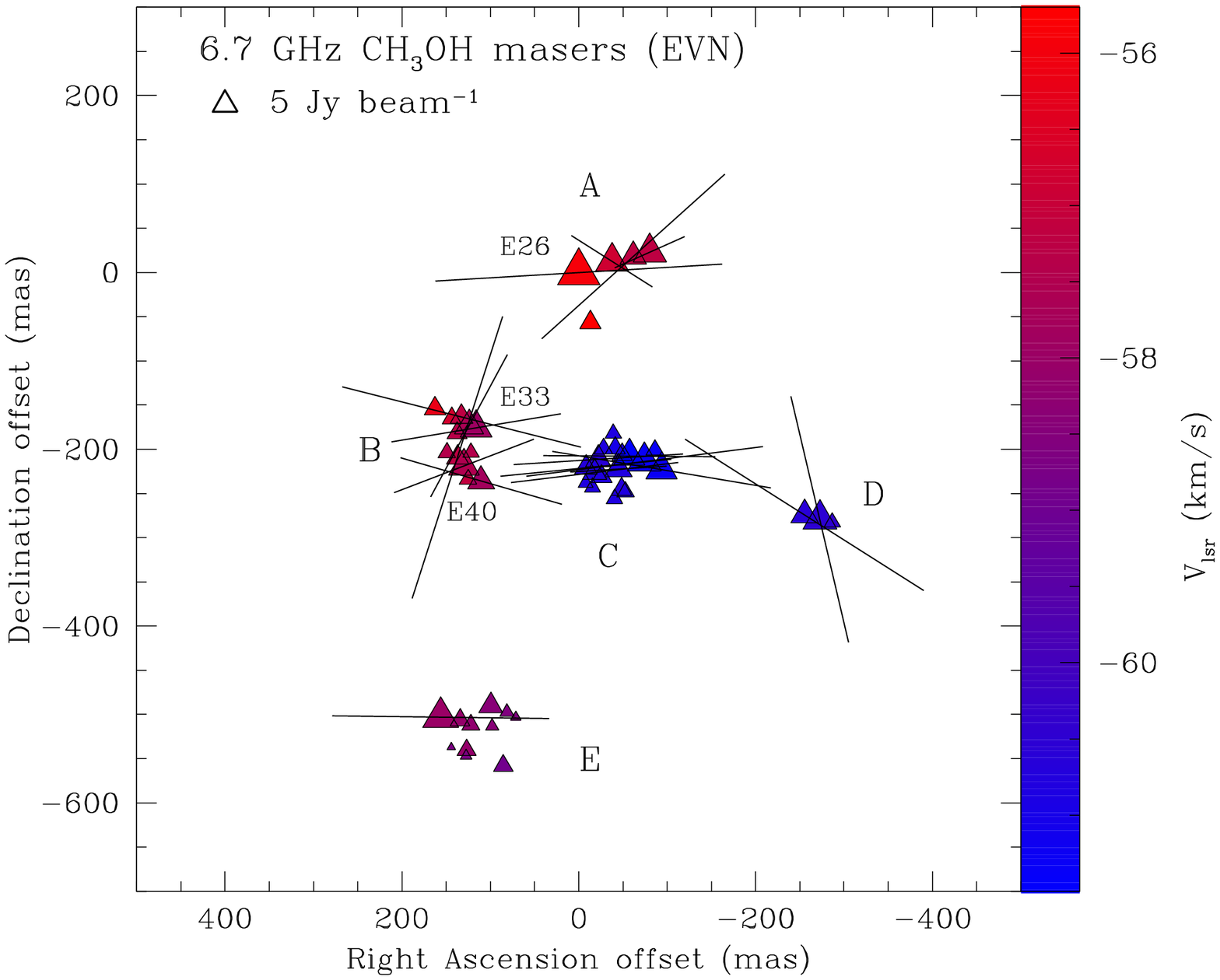}
\caption{Left panel: \meth ~maser structures detected with MERLIN around NGC7538-IRS1. Contours are 16, 32, 64, 128, 256, 512, 
1024~$\times$~0.15~\jyb. The synthesized 
beam is 47~mas~$\times$~34~mas. Right panel: a close-up view of the \meth ~maser features detected with the EVN around NGC7538-IRS1. 
The synthesized beam is 6.3~mas~$\times$~4.9~mas. The triangle symbols are the identified maser features in the present work scaled 
logarithmically according to their peak flux density. The maser LSR radial velocity is 
indicated by color. A 5~\jyb ~symbol is plotted for illustration in the right panels. 
The absolute reference position is $\alpha_{2000}=23^{\rm{h}}13^{\rm{m}}45^{\rm{s}}\!.362$ and
 $\delta_{2000}=+61^{\circ}28'10''\!\!.506$ as determined from the MERLIN observations.
The linear polarization 
vectors, scaled logarithmically according to the polarization fraction $P_{\rm{1}}$ (Tables~\ref{poltmer} and \ref{poltm}), are 
overplotted.}
\label{meth}
\end{figure*}
\begin{figure*}[th!]
\centering
\includegraphics[width = 7 cm]{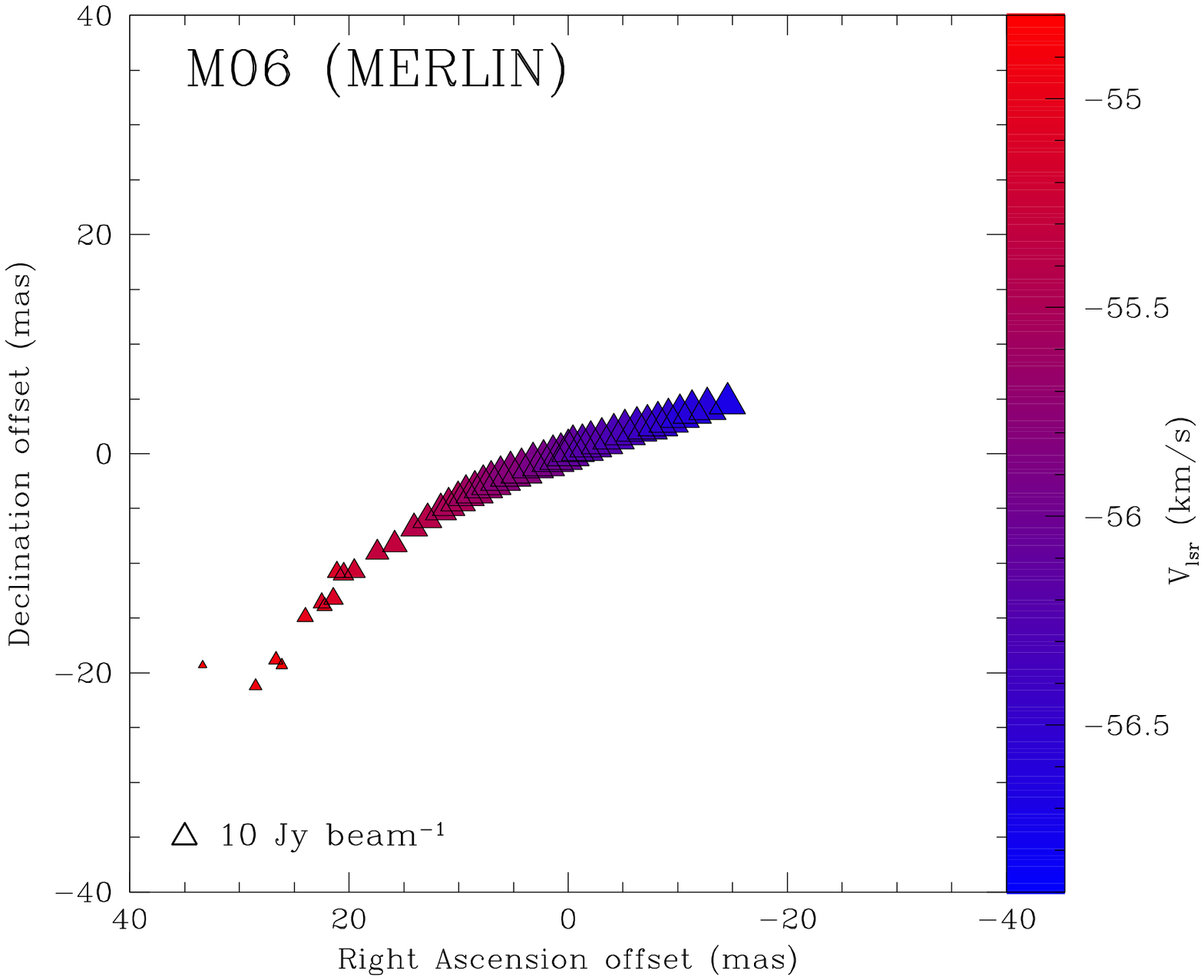}
\includegraphics[width = 7 cm]{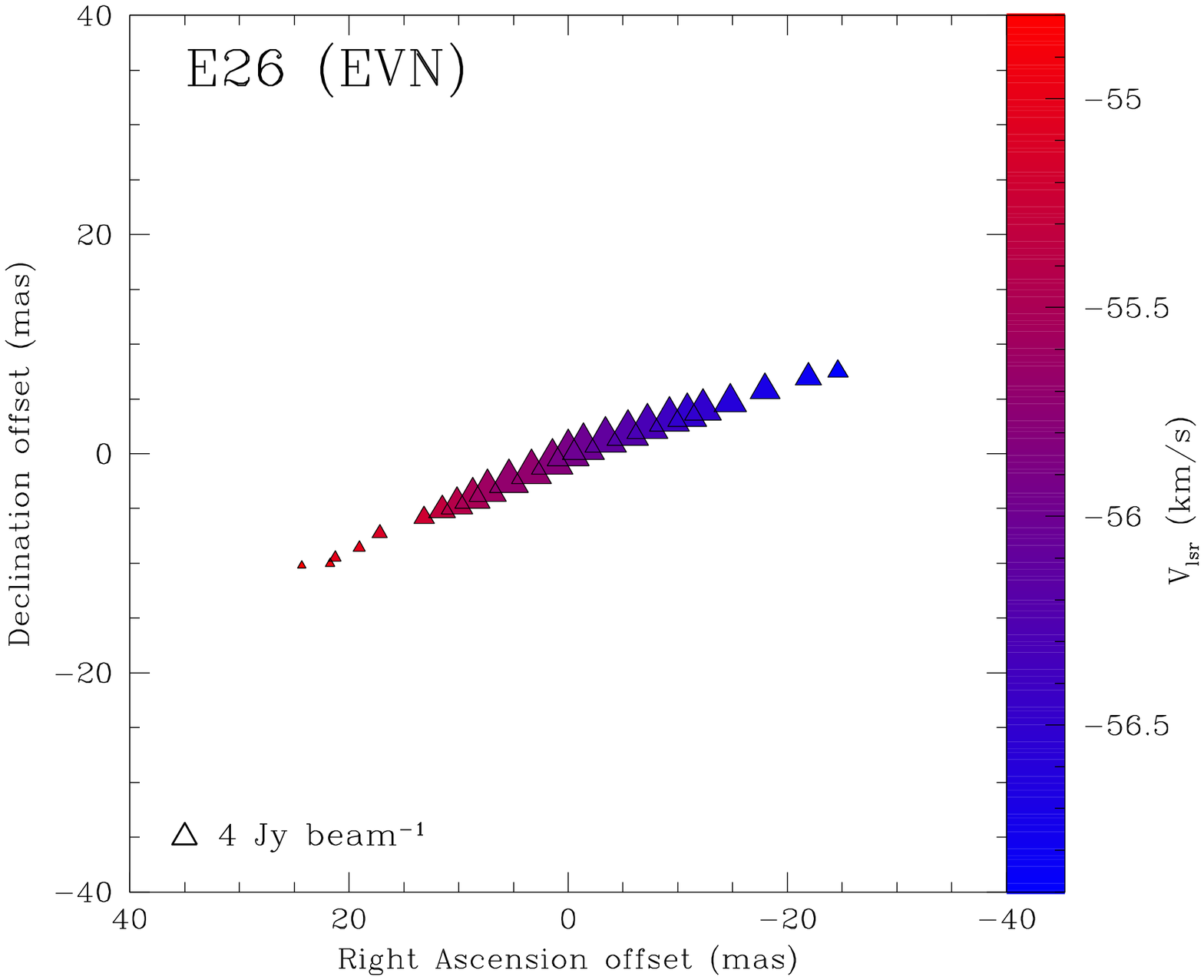}
\includegraphics[width = 7 cm]{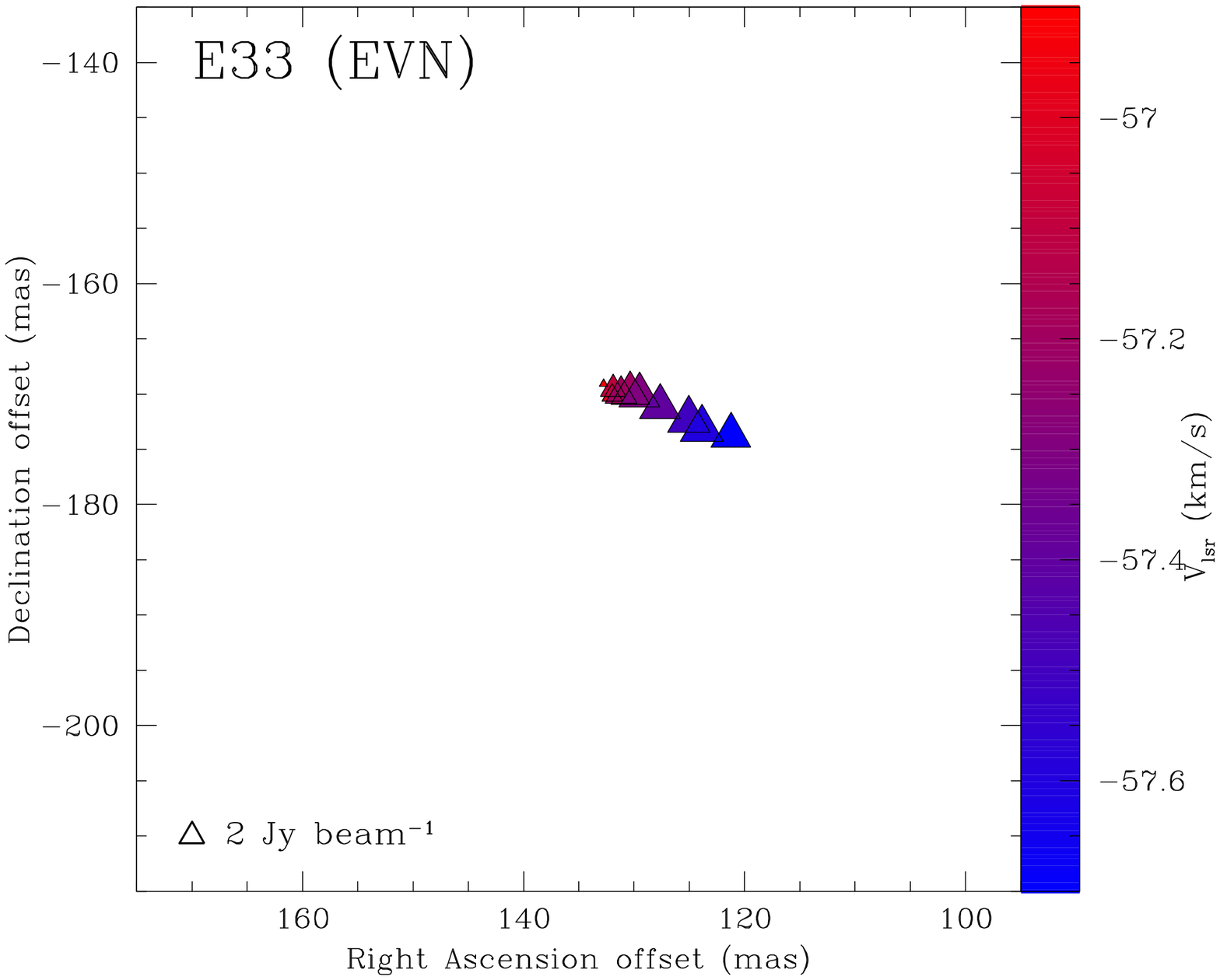}
\includegraphics[width = 7 cm]{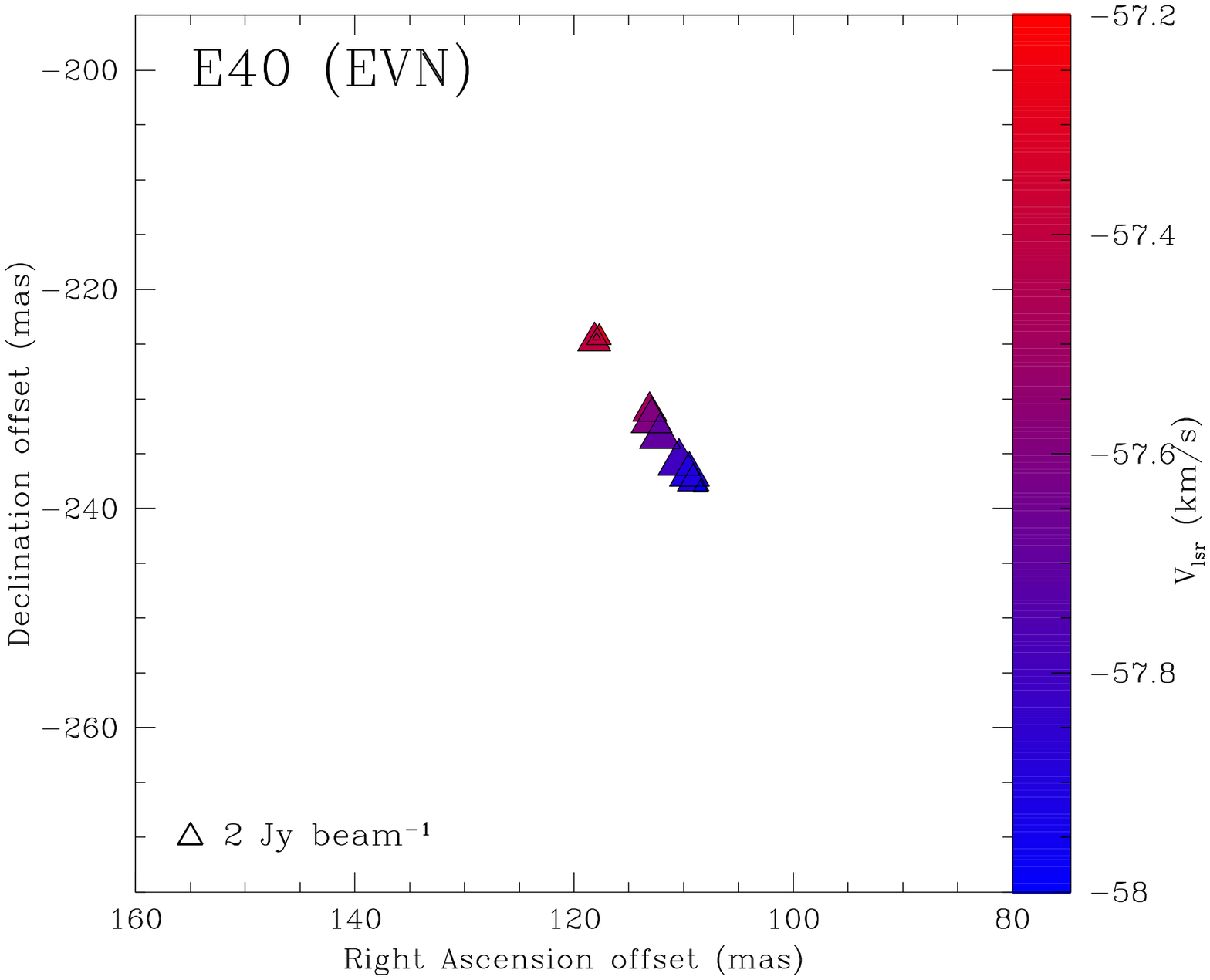}
\caption{Zoom-in view of the \meth ~maser spots of the features M06, which was detected with MERLIN, E26, E33, and E40, 
which were detected with the EVN. Features M06 and E26 are the same \meth ~maser detected with the two instruments. 
The LSR radial velocity of the spots is indicated by color. Ten, 4, and 2~\jyb ~symbols are plotted for illustration in the panels.}
\label{stream}
\end{figure*}
\begin {table*}[th!]
\caption []{All 6.7 GHz \meth ~maser features detected in NGC7538-IRS\,1 with the EVN.} 
\begin{center}
\scriptsize
\begin{tabular}{ l c c c c c c c c c c c c }
\hline
\hline
\,\,\,\,\,(1)&(2)& (3)   & (4)      & (5)          & (6)            & (7)                 & (8)          & (9)     & (10)                    & (11)                        & (12)                   & (13)    \\
\hline
Maser & group &RA        & Dec      & Peak flux    & $V_{\rm{lsr}}$ & $\Delta v\rm{_{L}}$ &$P_{\rm{l}}$ &  $\chi$  & $\Delta V_{\rm{i}}^{a}$ & $T_{\rm{b}}\Delta\Omega^{a}$& $\Delta V_{\rm{Z}}^{b}$& $\theta^{c}$\\
      &       &offset    & offset   & Density(I)   &                &                     &             &	   &                         &                             &                        &            \\ 
      &       & (mas)    & (mas)    & (\jyb)       &  (\kms)        &      (\kms)         & (\%)        &   (\d)   & (\kms)                  & (log K sr)                  & (m\,s$^{-1}$)          &  (\d)       \\
\hline
E01   &  D    & -286.910 & -282.502 &$0.55\pm0.02$ &  -60.40        &     0.12            &$-$          &  $-$     &  $-$                    & $-$                         & $-$	      	    & $-$ \\
E02   &  D    & -272.725 & -279.037 &$16.82\pm0.03$& -60.49         &     0.21            &$4.5\pm0.8$  &$+13\pm16$&  $1.0^{+0.3}_{-0.3}$    & $9.41^{+0.94}_{-0.39}$  	   & $+2.7\pm0.3$           & $77^{+13}_{-11}$ \\
E03$^{d}$&  D & -255.344 & -273.992 &$4.72\pm0.03$ & -60.58         &     0.24            &$5.6\pm0.9$  & $+58\pm5$&  $0.9^{+0.5}_{-0.1}$    & $9.58^{+0.81}_{-0.47}$  	   & $-$	            & $80^{+8}_{-10}$ \\
E04   &  C    & -93.660  & -222.357 &$10.39\pm0.02$& -61.28         &    0.24             &$3.6\pm0.2$  &$+81\pm22$&  $1.0^{+0.3}_{-0.2}$    & $9.26^{+0.58}_{-0.14}$  	   & $-$	            & $83^{+7}_{-23}$ \\
E05   &  C    & -86.171  & -200.790 &$0.73\pm0.02$ & -61.28         &     0.20            &$-$          &  $-$     &  $-$                    & $-$                     	   & $-$	            & $-$ \\
E06   &  A    & -80.377  & 22.786   &$15.21\pm0.03$&-57.24          &     0.20            &$3.6\pm0.2$  &$-66\pm10$&  $1.0^{+0.2}_{-0.2}$    & $8.74^{+0.69}_{-0.11}$  	   & $-$	            & $87^{+8}_{-17}$ \\
E07   &  C    & -74.392  & -213.730 &$16.41\pm0.03$& -61.28         &     0.29            &$4.1\pm0.5$  &$-83\pm20$&  $1.4^{+0.3}_{-0.5}$    & $9.39^{+0.92}_{-0.25}$ 	   & $-$ 	            & $77^{+8}_{-42}$ \\
E08   &  A    & -61.547  & 18.163   &$4.08\pm0.02$ & -57.15         &     0.23            &$4.3\pm1.4$  &$-48\pm40$&  $0.8^{+0.2}_{-0.3}$    & $9.37^{+0.61}_{-1.64}$  	   & $-$ 	            & $79^{+11}_{-22}$ \\
E09   &  C    & -57.174  & -207.658 &$9.48\pm0.02$ &  -61.46        &     0.27            &$2.5\pm0.7$  &$-90\pm4$ &  $1.3^{+0.3}_{-0.4}$    & $9.10^{+0.94}_{-1.25}$      & $-$                    & $75^{+15}_{-36}$ \\
E10   &  C    & -52.938  & -248.180 &$0.73\pm0.02$ &  -60.84        &     0.31            & $-$         &  $-$     &  $-$                    & $-$                         & $-$	            & $-$ \\
E11   &  C    & -49.085  & -205.700 &$1.12\pm0.03$ &  -60.75        &     1.08            &$-$          &  $-$     &  $-$                    & $-$                         & $-$	            & $-$ \\
E12   &  C    & -48.483  & -245.093 &$1.21\pm0.02$ &  -60.67        &     0.28            &$-$          &  $-$     &  $-$                    & $-$                         & $-$	            & $-$ \\
E13   &  C    & -44.930  & -222.357 &$4.65\pm0.02$ &  -60.84        &     0.25            &$1.4\pm0.4$  &$-86\pm83$&  $1.2^{+0.3}_{-0.3}$    & $8.81^{+0.83}_{-1.00}$      & $-$      	            & $76^{+14}_{-38}$ \\
E14   &  C    & -41.350  & -198.801 &$1.11\pm0.01$ &  -61.02        &     0.49            &$-$          &  $-$     &  $-$                    & $-$                         & $-$	            & $-$ \\
E15   &  C    & -40.421  & -255.891 &$0.59\pm0.02$ &  -60.84        &     0.33            &$-$          &  $-$     &  $-$                    & $-$                         & $-$	            & $-$ \\
E16   &  C    & -39.082  & -181.780 &$0.57\pm0.03$ &  -60.75        &     0.18            &$-$          &  $-$     &  $-$                    & $-$                         & $-$	            & $-$ \\
E17  &  A    & -37.661  & 12.650   &$14.31\pm0.02$&-56.80          &     0.33            &$1.4\pm0.2$  &$+58\pm12$& $1.6^{+0.3}_{-0.4}$     & $8.77^{+0.75}_{-0.25}$      & $-$     	            & $80^{+10}_{-40}$ \\
E18   &  C    & -27.986  & -199.362 &$0.80\pm0.03$ &  -60.75        &     0.19            &$-$          &  $-$     &  $-$                    & $-$                         & $-$ 	            & $-$ \\
E19   &  C    & -25.963  & -230.701 &$1.29\pm0.02$ &  -60.84        &     0.26            &$-$          &  $-$     &  $-$                    & $-$                         & $-$	            & $-$ \\
E20   &  C    & -22.083  & -211.354 &$3.19\pm0.03$ &  -60.58        &     0.27            &$2.4\pm1.0$  &$-87\pm23$&  $1.0^{+0.3}_{-0.4}$    & $9.05^{+0.39}_{-1.86}$      & $-$ 	            & $79^{+11}_{-42}$ \\
E21   &  C    & -18.147  & -226.401 &$2.53\pm0.03$ &  -60.75        &     0.22            &$2.41\pm0.04$&$-83\pm12$&  $1.0^{+0.3}_{-0.2}$    & $9.05^{+0.56}_{-0.06}$      & $-$	            & $85^{+6}_{-41}$ \\
E22   &  C    & -15.769  & -242.950 &$0.51\pm0.02$ &  -60.84        &     0.25            &$-$          &  $-$     &  $-$                    & $-$                         & $-$	            & $-$ \\
E23   &  A    & -13.091  & -56.523  &$1.43\pm0.05$ & -55.75         &     0.26            &$-$          &  $-$     &  $-$                    & $-$                         & $-$                    & $-$ \\
E24   &  C    & -8.172   & -220.826 &$1.83\pm0.02$ &  -60.67        &     0.29            &$2.5\pm0.3$  &  $-84\pm14$&  $1.3^{+0.3}_{-0.4}$  & $9.09^{+0.67}_{-0.22}$      & $-$	            & $82^{+7}_{-39}$ \\
E25   &  C    & -7.762   & -238.071 &$0.42\pm0.02$ &  -60.84        &     0.32            &$-$          &  $-$     &  $-$                    & $-$                         & $-$	            & $-$ \\
E26$^{d}$&  A & 0        & 0        &$95.15\pm0.08$&-55.92          &     0.33            &$5.83\pm0.03$&$-87\pm17$&  $0.5^{+0.4}_{-0.2}$    & $11.39^{+0.75}_{-0.03}$     & $+1.6\pm0.3$           & $86^{+5}_{-19}$\\
E27   &  E    & 71.113   & -503.039 &$0.19\pm0.01$ &  -58.56        &     0.12            &$-$          &  $-$     &  $-$                    & $-$                         & $-$	            & $-$ \\
E28   &  E    & 81.252   & -497.234 &$0.37\pm0.01$ &  -58.56        &     0.27            &$-$          &  $-$     &  $-$                    & $-$                         & $-$	            & $-$ \\
E29   &  E    & 85.297   & -557.914 &$1.06\pm0.01$ &  -58.91        &     0.27            &$-$          &  $-$     &  $-$                    & $-$                         & $-$	            & $-$ \\
E30   &  E    & 97.923   & -513.329 &$0.30\pm0.01$ &  -58.38        &     0.18            &$-$          &  $-$     &  $-$                    & $-$                         & $-$	            & $-$ \\
E31   &  E    & 99.399   & -489.943 &$2.26\pm0.01$ &  -58.38        &     0.21            &$-$          &  $-$     &  $-$                    & $-$                         & $-$	            & $-$ \\
E32   &  B    & 110.386  & -235.856 &$4.37\pm0.02$ &  -57.77        &     0.31            &$2.4\pm0.6$  &$+74\pm44$& $1.4^{+0.3}_{-0.4}$     & $9.06^{+0.81}_{-1.33}$      & $-$                    & $89^{+10}_{-38}$ \\
E33   &  B    & 116.016  & -175.760 &$10.11\pm0.02$& -57.86         &     0.30            &$2.5\pm0.7$  &$-81\pm83$& $1.5^{+0.4}_{-0.4}$     & $9.05^{+0.94}_{-1.28}$      & $-$	            & $76^{+15}_{-36}$ \\
E34   &  B    & 121.782  & -203.613 &$0.54\pm0.02$ &  -57.42        &     0.55            &$-$          &  $-$     &  $-$                    & $-$                         & $-$	            & $-$ \\
E35   &  E    & 122.083  & -511.884 &$0.80\pm0.02$ &  -58.21        &     0.22            &$-$          &  $-$     &  $-$                    & $-$                         & $-$	            & $-$ \\
E36   &  B    & 123.859  & -173.165 &$6.46\pm0.02$ &  -57.59        &     0.32            &$2.3\pm0.6$  &$-28\pm42$&  $1.5^{+0.4}_{-0.4}$    & $9.04^{+1.00}_{-1.08}$      & $-$	            & $89^{+12}_{-38}$\\
E37   &  B    & 125.062  & -233.976 &$0.68\pm0.02$ &  -57.07        &     0.28            &$-$          &  $-$     &  $-$                    & $-$                         & $-$	            & $-$ \\
E38   &  E    & 126.893  & -540.110 &$0.93\pm0.03$ &  -58.12        &     0.21            &$-$          &  $-$     &  $-$                    & $-$                         & $-$	            & $-$ \\
E39   &  E    & 127.303  & -546.631 &$0.21\pm0.06$ &  -58.73        &     0.26            &$-$          &  $-$     &  $-$                    & $-$                         & $-$	            & $-$ \\
E40   &  B    & 129.927  & -218.961 &$8.98\pm0.02$ &  -57.51        &     0.37            &$2.1\pm0.7$  &$-69\pm10$&  $1.8^{+0.1}_{-0.4}$    & $8.99^{+1.00}_{-1.5}$       & $-$  	            & $71^{+9}_{-44}$ \\
E41$^{d}$&  B & 132.550  & -163.392 &$1.81\pm0.02$ &  -57.24        &     0.34            &$4.3\pm1.1$  &$+76\pm42$& $0.6^{+0.5}_{-0.1}$     & $9.44^{+0.50}_{-1.72}$      & $-$	            & $85^{+5}_{-17}$ \\
E42   &  E    & 134.190  & -506.261 &$1.10\pm0.03$ &  -58.12        &     0.20            &$-$          &  $-$     &  $-$                    & $-$                         & $-$     	            & $-$ \\
E43$^{d}$&  B & 137.196  & -209.274 &$1.69\pm0.03$ &  -57.24        &     0.37            &$6.2\pm2.0$  &$-18\pm17$&  $0.5^{+0.3}_{-0.3}$    & $9.83^{+0.22}_{-1.94}$      & $-$     	            & $90^{+20}_{-20}$\\
E44   &  B    & 137.552  & -181.682 &$1.05\pm0.02$ &  -57.33        &     0.24            &$-$          &  $-$     &  $-$                    & $-$                         & $-$	            & $-$ \\
E45   &  B    & 143.509  & -164.852 &$0.98\pm0.02$ &  -56.72        &     0.32            &$-$          &  $-$     &  $-$                    & $-$                         & $-$	            & $-$ \\
E46   &  E    & 144.083  & -537.047 &$0.12\pm0.01$ &  -58.73        &     0.28            &$-$          &  $-$     &  $-$                    & $-$                         & $-$	            & $-$ \\
E47  &  B    & 149.112  & -203.584 &$0.71\pm0.02$ &  -57.42        &     0.27            &$-$          &  $-$     &  $-$                    & $-$                         & $-$	            & $-$ \\
E48   &  E    & 156.163  & -503.223 &$23.68\pm0.03$& -58.03         &     0.24            &$3.5\pm0.6$  &$+89\pm44$&  $1.2^{+0.2}_{-0.4}$    & $9.26^{+1.00}_{-0.31}$      & $-2.7\pm0.3$           & $75^{+15}_{-35}$ \\
E49   &  B    & 162.558  & -154.095 &$1.50\pm0.05$ &  -56.19        &     0.16            &$-$          &  $-$     &  $-$                    & $-$                         & $-$    	            & $-$ \\
\hline
\end{tabular}
\end{center}
\scriptsize{\textbf{Notes.} 
$^{(a)}$ The best-fitting results obtained by using a model based on the radiative transfer theory of \meth ~masers 
for $\Gamma+\Gamma_{\nu}=1$s$^{-1}$ (Vlemmings et al. \cite{vle10}). The errors were determined by analyzing the full probability 
distribution function. For $\Gamma+\Gamma_{\nu}=0.6$s$^{-1}$ (Minier et al. \cite{min02}) $T_{\rm{b}}\Delta\Omega$ has to be adjusted by adding 
$-0.22$. 
$^{(b)}$ The Zeeman-splittings are determined from the cross-correlation between the RR and LL spectra. 
$^{(c)}$The angle between the magnetic field and the maser propagation direction is determined by using the observed $P_{\rm{l}}$ 
and the fitted emerging brightness temperature. The errors were determined by analyzing the full probability distribution function.
$^{(d)}$Because of the degree of the saturation of these \hdueo ~masers, $T_{\rm{b}}\Delta\Omega$ is underestimated, $\Delta V_{\rm{i}}$ 
and $\theta$ are overestimated.}
\label{poltm}
\end{table*}
\subsection{\meth~ masers}
We detected 13 6.7-GHz \meth ~maser features with the MERLIN telescope (named M01-M13 in Table~\ref{poltmer}) that appear to be
composed of 49 features when observed with the EVN resolution (named E01-E49 in Table~\ref{poltm}). Including the more
sensitive EVN observations, we detect no \meth ~maser emission below $0.1$~\jyb. Note that as defined in Sect.~\ref{obssect}, we indicate 
with the term maser feature the brightest maser spot among a series of maser spots that either show a spatial
 coincidence and consecutive velocities or show a clear velocity gradient along a continuum linear structure.\\
\indent In the left panel of Fig.~\ref{meth} we show the contours of the \meth ~maser structures detected with MERLIN and in the 
right panel the distribution of the \meth ~maser features detected with the EVN. The \meth ~maser features distribution at 
high angular resolution match the \meth ~maser emission detected with MERLIN four years before perfectly.
Following the naming convention adopted by Minier et al. (\cite{min00}), they can be divided into five groups (from A 
to E). 
Note that each group is composed of several maser features, each of which indicates a series of maser spots.
Group A, which is composed of five maser features (3 at MERLIN resolution) and hosts the brightest maser feature of the region 
(M06 and E26, respectively), shows a linear distribution. If we consider all maser spots of 
the maser feature M06 (top-left panel of Fig.~\ref{stream}), we are able to measure a velocity gradient of about 
0.02~\kms~AU$^{-1}$ from northwest to southeast, which is confirmed by considering the matching maser spots of E26 (EVN, top-right 
panel of Fig.~\ref{stream}).\\
\indent Group B is resolved into 12 \meth ~maser features (four at MERLIN resolution). The spots of the features E33 and 
E40 of this group show a velocity gradient similar to that of E26 but from southwest to northeast (bottom panels of 
Fig.~\ref{stream}). No \meth ~maser features of group B detected with MERLIN show a velocity gradient. The \meth ~maser groups C, D, 
and E are composed of 18, 3, and 11 maser features, respectively. While groups C and D are located close to the central peak of the 
continuum emission, group E is located about 300~mas southward. The velocities of groups C and D are more blue-shifted 
($-61.5$~\kms~$<V_{\rm{lsr}}<-60.5$~\kms) than those of groups A, B, and E ($-59$~\kms~$<V_{\rm{lsr}}<-56$~\kms).\\
\indent We detected linear polarization in 10 and 20 \meth ~maser features with MERLIN and EVN, respectively. The features of 
group D show the highest linear polarization fraction of the region (column 8 of Tables~\ref{poltmer} and \ref{poltm}), though 
the observations at higher angular resolution revealed E43 to be the feature with the highest linear polarization fraction 
(6.2\%). Because the EVN provides an angular resolution eight times better than MERLIN and because we also have a higher dynamic range, we will,
 for the interpretation of the magnetic field, only use the linear polarization vectors of the 
maser features detected with the EVN. The groups A, C, and E have weighted linear polarization vectors 
almost oriented east-west, with angles $-86$\d$\pm31$\d, $-88$\d$\pm6$\d, and $+89$\d$\pm44$\d, respectively. The weighted 
linear polarization angles for the other two groups are $\langle\chi_{\rm{CH_3OH}}^{\rm{B}}\rangle=-60$\d$\pm36$\d ~and 
$\langle\chi_{\rm{CH_3OH}}^{\rm{D}}\rangle=+48$\d$\pm32$\d.\\
\begin{figure}[h!]
\centering
\includegraphics[width = 8.8 cm]{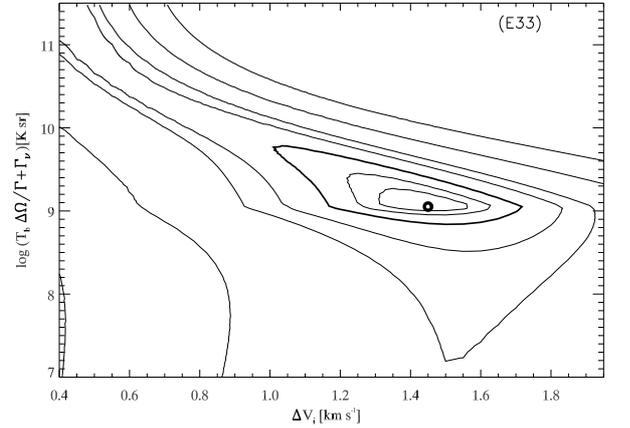}
\caption{Result of the full radiative transfer $\chi^{2}$-model fits for the \meth ~maser E33 detected with the EVN. The fit 
yields the emerging maser brightness temperature $T_{\rm{b}}\Delta\Omega$ and the intrinsic maser thermal linewidth $\Delta V_{\rm{i}}$. 
Contours indicate the significance intervals $\Delta \chi^{2}$=0.25, 0.5, 1, 2, 3, 7, with the thick solid contours indicating 
1$\sigma$ and 3$\sigma$ areas.}
\label{evnfit}
\end{figure}
\indent The full radiative transfer method code for \meth ~masers was able to fit 7 (MERLIN) and 20 (EVN) \meth ~maser features, 
the results are given in columns 10 and 11 of Tables~\ref{poltmer} and \ref{poltm}. Considering the high angular resolution 
observations, the weighted intrinsic maser linewidth and the weighted emerging brightness temperature are 
$\langle \Delta V_{\rm{i}}\rangle_{\rm{CH_3OH}}=0.9$~\kms ~and 
$\langle T_{\rm{b}}\Delta\Omega\rangle_{\rm{CH_3OH}}\approx10^{9}$~K~sr. As an example, the $\chi^{2}$-contours for feature E33 are 
reported In Fig.~\ref{evnfit}.  
\begin {table*}[th!]
\caption []{Weighted values of the linear polarization angles, the intrinsic thermal linewidths, the emerging brightness temperatures, 
and the angles between the line of sight and the magnetic field of the \meth ~masers for feature in common.} 
\begin{center}
\scriptsize
\begin{tabular}{ c c c c c c c c c c}
\hline
Group & \multicolumn{4}{c}{EVN} & & \multicolumn{4}{c}{MERLIN}\\
\cline{2-5} \cline{7-10} 
   & $\langle\chi\rangle$&$\langle \Delta V_{\rm{i}}\rangle$ & $\langle T_{\rm{b}}\Delta\Omega\rangle$ & $\langle \theta \rangle$  & &$\langle\chi\rangle$
&$\langle \Delta V_{\rm{i}}\rangle$ & $\langle T_{\rm{b}}\Delta\Omega\rangle$ &$\langle \theta \rangle$ \\
   & (\d) &(\kms) 			        & (K~sr)                                  &(\d)                       & &(\d)& (\kms) & (K~sr) &(\d) \\
\hline
 A & $-86\pm31$&$0.9^{+0.4}_{-0.3}$& $5\times10^{9}$ & $86^{+4}_{-42}$ & & $-51\pm62$ &$<$1.2$^{a}$        &$>5\times10^{8}$$^{b}$   & $90^{+53}_{-53}$     \\
 B & $-60\pm36$&$0.5^{+0.9}_{-0.1}$& $1\times10^{9}$ & $87^{+4}_{-42}$  & & $-63\pm76$ &$<$0.9$^{a}$        &$>1\times10^{12}$$^{b}$ & -  \\
 C & $-88\pm6$ &$1.0^{+0.3}_{-0.4}$& $2\times10^{9}$ & $85^{+4}_{-46}$ &  & $-74\pm69$ &$2.2^{+0.1}_{-0.1}$&$3\times10^{8}$          & $76^{+11}_{-46}$    \\
 D & $+48\pm32$&$0.6^{+0.4}_{-0.1}$& $2\times10^{10}$& $79^{+8}_{-13}$ & & $+39\pm8$  &$<$0.6$^{a}$        &$>1\times10^{12}$$^{b}$  & -       \\
 E & $+89\pm44$&$1.2^{+0.2}_{-0.4}$& $2\times10^{9}$ & $75^{+15}_{-35}$ & & $+21\pm54$ &$1.5^{+0.1}_{-0.1}$ &$6\times10^{8}$         & $86^{+5}_{-29}$  \\  
\hline
\end{tabular}
\end{center}
\scriptsize{\textbf{Notes.} 
$^{(a)}$ Only the highest value has been taken into account.
$^{(b)}$ Only the lowest value has been taken into account.}
\label{weight}
\end{table*}

\indent The fit for the \meth ~maser features detected with 
MERLIN gives values both for $\Delta V_{\rm{i}}$ and for $T_{\rm{b}}\Delta\Omega$ consistent or higher than
those detected with the EVN (see Table~\ref{weight}). However, owing to the lower angular resolution of MERLIN and the occurrence of strong 
velocity gradients, the fits are strongly affected by line blending. As for the \hdueo ~maser features, 
we are able to determine the $\theta$ values  (column 12 and 13 of Tables~\ref{poltmer} and \ref{poltm}, respectively) from 
$T_{\rm{b}}\Delta\Omega$ and $P_{\rm{l}}$. The weighted value for the whole region is 
$\langle \theta_{\rm{CH_3OH}} \rangle_{\rm{EVN}}=86$\d$^{+4^{\circ}}_{-43^{\circ}}$, which is almost constant in all groups. 
Only for group E we determine a lower value, i.e., 
$\langle \theta_{\rm{CH_3OH}}^{\rm{E}}\rangle_{\rm{EVN}}=75$\d$^{+15^{\circ}}_{-35^{\circ}}$.\\
\indent The Zeeman-splitting ($\Delta V_{\rm{Z}}$) in \ms ~determined from the cross-correlation between the RR and LL spectra of the 
\meth ~maser features detected with the EVN is reported in column 12 of Table~\ref{poltm}. Note that the cross-correlation method is 
dynamic-range sensitive, therefore we are able to measure $\Delta V_{\rm{Z}}$ only for those features that show a very high dynamic range. 
It has been impossible to obtain $\Delta V_{\rm{Z}}$ measurements from MERLIN data because of the insufficient dynamic range.
\section{Discussion}
\label{dis}
%
\subsection{Comparing \meth ~spectra at different resolution}
\label{comp}
\indent Vlemmings (\cite{vle08}) observed the \meth ~masers with the 100-m Effelsberg telescope 
in November 2007, i.e. two years later
than our MERLIN observations and two years before our EVN observations. A comparison between the single-dish and MERLIN fluxes reveals 
only a difference of $\lesssim$~10\% across the whole spectra. This slight difference can be explained by the different resolution of the two 
instruments and flux calibration uncertainties in the Effelsberg observations ($\sim$~10\%, Vlemmings \cite{vle08}).
This means that no significant changes occurred between the two epochs and there is no indication of a core/halo structure of the maser features
 (hereafter simply called masers) at scales of
$\sim$100~AU such as described for a sample of \meth ~maser sources by Pandian et al. (\cite{pan11}).\\
\begin{figure}[h!]
\centering
\includegraphics[width = 8 cm]{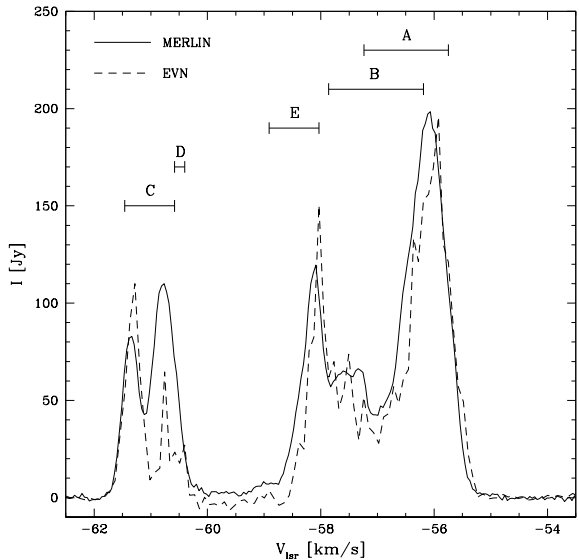}
\caption{Profile of the total flux spectra of the \meth ~masers in NGC7538 detected with MERLIN (solid line, epoch 2005) and
the EVN (dashed line, epoch 2009). The groups are also reported according to the velocity of Table~\ref{poltm}.}
\label{over}
\end{figure}
\indent In Fig.~\ref{over} we show the total flux spectra of the \meth ~masers detected with MERLIN and the EVN in 2005 and 2009, 
respectively. Generally, the EVN resolved out between 60\% and 90\% of the flux when 6.7-GHz \meth ~masers were observed with both these 
instruments (e.g., Minier et al. \cite{min02}; Bartkiewicz et al. \cite{bar09}; Torstensson et al. \cite{tor11}). This 
indicates the presence of emission structures on scales that are resolved by the EVN, i.e. a halo structure that surrounds the core of the masers.
 Moreover, if the maser spots do not have a compact core, they would not be detected in the EVN observations 
(Pandian et al. \cite{pan11}). Although in our observations the EVN resolves out about 50\% of the flux of group C ($V_{\rm{lsr}}>-61$~\kms),
 the majority of the flux appears to be recovered across the entire \meth ~maser spectrum. This suggests that the halo structure is absent, 
except in some masers of group C and likely in E26 of group A, and that the masers only contain a compact core.
Moreover, the non-detection of the maser M07 with the EVN might indicate the absence of the compact core in this maser.\\
\indent Minier et al. (\cite{min02}) observed the 6.7-GHz \meth ~masers with five antennas of the EVN. The authors determined the size of the 
halo ($d_{\rm{1}}$) and of the core ($d_{\rm{2}}$) of the brightest masers of groups A and C by Gaussian fitting of the visibility amplitudes. 
For the brightest maser of group A the authors determined $d_{\rm{1}}=5$~mas ($\sim$13~AU at 2.65~kpc) and $d_{\rm{2}}=3$~mas ($\sim$8~AU), and 
$d_{\rm{1}}=17$~mas ($\sim$45~AU), and $d_{\rm{2}}=5$~mas ($\sim$13~AU) were measured for the brightest maser of group C. The Gaussian fit of E26 
and E07 gives a size of 5~mas and 6~mas respectively, indicating that the EVN resolves out the halo structure observed by Minier et al. 
(\cite{min02}) and that we observe only the core of these two masers.\\
\indent We suggest that the absence of a halo structure in most of the \meth ~masers could be caused by the amplification of the 
strong continuum emission shown in Fig.~\ref{pos}. This would argue that almost all \meth ~masers are on the front-side of this
source. Similar arguments were used in the interpretation of the structure of Cepheus~A
(Torstensson et al. \cite{tor11}).
\subsection{\hdueo ~and \meth ~maser properties}
Before discussing the polarization analysis of the masers in NGC7538-IRS1, we need to consider the degree of the maser saturation. 
As explained
in detail by Vlemmings et al. (\cite{vle10}) and Surcis et al. (\cite{sur11}), when the full radiative transfer method code is applied
 to a saturated maser, the code gives a lower limit for $T_{\rm{b}}\Delta\Omega$ and an upper limit for $\Delta V_{\rm{i}}$.
The masers are unsaturated when $R/(\Gamma+\Gamma_{\nu})<1$ and fully saturated when $R/(\Gamma+\Gamma_{\nu})\sim100$, where R 
is the stimulated emission rate given by
\begin{equation}
 R\simeq\frac{Ak_{\rm{B}}T_{\rm{b}}\Delta\Omega}{4\pi h\nu}.
\end{equation}
Here $A$ is the Einstein coefficient for the maser transition, which for \hdueo ~masers is taken to be equal to $2 \times10^{−9} ~\rm{s}^{-1}$ 
(Goldreich \& Keeley \cite{gol72}) and for the \meth ~masers is $0.1532 \times10^{−8} ~\rm{s}^{-1}$ (Vlemmings et al. \cite{vle10}),
$k_{\rm{B}}$ and $h$ are the Boltzmann and Planck constants, respectively, and $\nu$ the maser frequency. From $R/(\Gamma+\Gamma_{\nu})<1$
we can estimate an upper limit for the emerging brightness temperature below which the masers can be considered unsaturated. 
The limits are  $(T_{\rm{b}}\Delta\Omega)_{\rm{H_2O}}<6.7\times10^{9}$~K~sr and 
$(T_{\rm{b}}\Delta\Omega)_{\rm{CH_3OH}}<2.6\times10^{9}$~K~sr for \hdueo ~and \meth ~masers, respectively. Consequently, only four \meth 
~masers are partly saturated; i.e., E03, E26, E41 and E43 (Table~\ref{poltm}). Because the model emerging brightness temperature
 scales linearly with $\Gamma+\Gamma_{\nu}$, the ratio $R/(\Gamma+\Gamma_{\nu})$ is independent of the value of $\Gamma+\Gamma_{\nu}$. 
The $\Delta v\rm{_{L}}$ of the saturated \meth ~masers are close to $\Delta V_{\rm{i}}$ implying 
that the maser lines are rebroadened as expected when the maser becomes saturated. Furthermore, these four \meth ~masers do also show a high
 linear polarization fraction, which again confirms their saturated state (Goldreich et al. \cite{gol73}).\\
\indent The intrinsic linewidth of the \hdueo ~maser W07 is higher than the typical values measured in previous works, i.e. 
$\Delta V_{\rm{i}}=2.5$~\kms ~(Surcis et al. \cite{sur11}). This difference might indicate the presence
of a strong turbulent gas with a turbulence velocity of $\Delta V_{\rm{turb}}=2.3$~\kms ~or multiple components overlapping.\\
\indent By comparing the brightness temperature $T_{\rm{b}}$ with $T_{\rm{b}}\Delta\Omega$ obtained from the model, we can  
estimate the maser beaming angle ($\Delta\Omega$) for both maser species. We can estimate the brightness temperature ($T_{\rm{b}}$) by 
considering the equation
\begin{equation}
 \frac{T_{\rm{b}}}{[\rm{K}]}=\frac{S(\nu)}{[\rm{Jy}]}\cdot\left(\frac{\Sigma^{2}}{[\rm{mas}^2]}\right)^{-1} \cdot \xi,
\end{equation}
where $S(\nu)$ is the flux density, $\Sigma$ the maser angular size and $\xi$ is a constant factor that includes all constant values,
such as the Boltzmann constant, the wavelength, and the proportionality factor obtained for a Gaussian shape by Burns et al. 
(\cite{bur79}). The values of $\xi$ for \hdueo ~and \meth ~masers are
\begin{equation}
 \xi_{22\,\rm{GHz}}=1.24 \cdot 10^{9} \rm{~mas^{2} ~Jy^{-1} ~K},
\end{equation}
\begin{equation}
\xi_{6.7\,\rm{GHz}}=13.63 \cdot 10^{9} \rm{~mas^{2} ~Jy^{-1} ~K},
\end{equation}
as reported by Surcis et al. (\cite{sur11}) and Surcis et al. (\cite{sur09}), respectively. The \hdueo ~maser W07 is unresolved 
and $\Delta\Omega_{\rm{H_2O}}=10^{-2}$. The Gaussian fit
of \meth ~masers detected with the EVN gives a size between 5 and 7 mas, which indicates that all \meth ~masers are 
marginally resolved.\\
\indent In a tubular geometry $\Delta\Omega\approx(d/l)^{2}$, where $d$ and $l$ are the transverse size and length of the tube, 
respectively. By assuming $d$ approximately the size of the masers, the maser lengths are in the range 
$10^{13}$~cm~$<l_{\rm{H_2O}}<10^{14}$~cm and $10^{14}$~cm~$<l_{\rm{CH_3OH}}<10^{15}$~cm for W07 and the \meth ~masers, respectively.
\subsection{Magnetic field in NGC7538-IRS\,1}
\subsubsection{\meth ~Zeeman-splitting and magnetic field}
\label{mfs}
We measured 6.7-GHz \meth ~maser Zeeman-splitting ranging from -2.7~\ms ~to +2.7~\ms. This is 2--3 times larger than was measured with the
Effelsberg telescope (Vlemmings \cite{vle08}) as expected when resolving individual masers. From the Zeeman theory we know that the 
Zeeman-splitting is related to the magnetic field strength along the line of sight ($B_{||}$) by the equation
\begin{equation}
\Delta V_{\rm{Z}}= \alpha_{\rm{Z}} \cdot B_{||},
\label{zem}
\end{equation}
where $\Delta V_{\rm{Z}}$ is the Zeeman-splitting in \kms ~and $\alpha_{\rm{Z}}$ is the Zeeman-splitting coefficient  in 
\kmsg, which strongly depends on the Land\'{e} \textit{g}-factor of the molecular transition. Recently, Vlemmings et al. 
(\cite{vle11}) found an unfortunate calculation error in the $\alpha_{\rm{Z}}$ used so far ($\alpha_Z=0.049$~\kmsg; e.g., Surcis et al. 
\cite{sur09}, Vlemmings et al. 
\cite{vle10}), the current best value is found to be one order of magnitude lower ($\alpha_Z=0.005$~\kmsg). 
However, Vlemmings et al. (\cite{vle11}) show that the splitting between the RCP- and LCP-signal still likely originates from the Zeeman theory 
and not from other non-Zeeman effects. Until careful laboratory measurements of the \textit{g}-factor appropriate for the 6.7-GHz 
\meth ~masers are made, we cannot give any exact value for the magnetic field strength. Because there is a linear proportionality between 
$\Delta V_{\rm{Z}}$ and $B_{||}$, we can say that the negative value of the Zeeman-splitting indicates a magnetic field oriented toward 
the observer and positive away from the observer. 
However, we can speculatively give a possible range of values for $|B_{||}|$  by considering $0.005$~\kmsg$<\alpha_Z<0.049$~\kmsg,
thus $50$~mG~$\lesssim |B_{||}|\lesssim500$~mG.
\subsubsection{Magnetic field orientation}
\label{mfo}
Before discussing the orientation of the magnetic field, we have to evaluate the foreground Faraday rotation (i.e., 
the rotation caused by the medium between the source and the observer), which is given by
\begin{equation}
\Phi_{\rm{f}}[^{\circ}]=4.22\times10^{6}~\left(\frac{D}{[\rm{kpc}]}\right)~
\left(\frac{n_{e}}{[\rm{cm^{-3}}]}\right)~\left(\frac{B_{||}}{[\rm{mG}]}\right)~
\left(\frac{\nu}{[\rm{GHz}]}\right)^{-2},
\label{fari}
\end{equation}
\noindent where $D$ is the length of the path over which the Faraday rotation occurs, $n_{\rm{e}}$ and $B_{||}$ are respectively the 
average electron density and the magnetic field along this path and $\nu$ is the frequency. By assuming the interstellar electron density, 
the magnetic field, and the distance are $n_{\rm{e}}\approx0.012\,\rm{cm^{-3}}$, $B_{||}\approx2\,\rm{\mu G}$ (Sun et al. \cite{sun08}),
and $D=2.65$~kpc, respectively,
 $\Phi_{\rm{f}}$ is estimated to be 0\d$\!\!.5$ at 22-GHz and 6\d$\!\!.0$ at 6.7-GHz. At both frequencies the foreground Faraday 
rotation is within the errors reported in Tables~\ref{poltw} and \ref{poltm} thus it should not affect our conclusions. \\
\indent Because all $\theta$ values are higher than the Van Vleck angle $\theta_{\rm{crit}}\sim55$\d ~(Goldreich et al. \cite{gol73}), 
the magnetic fields are inferred to be perpendicular to the linear polarization vectors. The magnetic field orientation derived from the \hdueo 
~masers is $\varphi_{\rm{H_2O}}=+79$\d, while the five groups of \meth ~masers show orientation angles
$\varphi_{\rm{CH_3OH}}^{A}=+4$\d, $\varphi_{\rm{CH_3OH}}^{B}=+30$\d, $\varphi_{\rm{CH_3OH}}^{C}=+2$\d, $\varphi_{\rm{CH_3OH}}^{D}=-42$\d, 
and $\varphi_{\rm{CH_3OH}}^{D}=-1$\d. However, considering the relatively large errors in column 13 of Table~\ref{poltm}, we cannot 
rule out that the actual $\theta$ values are below 55\d, in which case the magnetic field is parallel to the linear polarization 
vectors.\\
\indent Because Hutawarakorn \& Cohen (\cite{hut03}) measured the linear polarized emission of the 1.6 and 1.7-GHz OH masers, we can now 
compare the orientation of the magnetic field derived from the two maser species (Fig.~\ref{pos}). We have to consider that the Faraday rotation 
at the OH maser frequencies is, at the same conditions, larger than at 6.7-GHz. The linear polarization angles of group D show a 90\d 
~difference w.r.t those of the nearby 1.6-GHz OH masers. 
This might be caused either by a 90\d--flip phenomenon or by the Faraday rotation if the masers are deeply located in a strong ionized gas.
Most of the OH masers are likely to be Zeeman $\sigma$-components, therefore the magnetic field is perpendicular to the 
linear polarization vectors. Considering the large errors of $\theta_{\rm{E48}}$, the orientation of the magnetic fields of group B and E are 
consistent with those derived from the nearby 1.6-GHz OH masers (Hutawarakorn \& Cohen \cite{hut03}). In particular, the linear polarization 
vector of E43, which shows the highest linear polarization fraction, is perfectly aligned with feature 2 of the 1665~MHz OH maser 
(Hutawarakorn \& Cohen \cite{hut03}). Unless the Faraday rotation is 180\d, the consistence of the magnetic field orientation indicates that
the Faraday rotation at OH maser frequency is low.
%
\subsubsection{The role of the magnetic field}
\indent The importance of the magnetic field in the region can be estimated by evaluating the ratio between thermal and magnetic
energies ($\beta$). If $\beta<1$, the magnetic field dominates the energies in the high-density protostellar environment. The $\beta$
factor is given by
\begin{equation}
\beta=2 \left(\frac{m_{\rm{a}}}{m_{\rm{s}}}\right)^{2},
\label{beta}
\end{equation}
where $m_{\rm{a}}$ is the Alfv\'{e}nic Mach number and $m_{\rm{s}}$ is the sonic Mach number, which in formula are
\begin{equation}
m_{\rm{a}}=\frac{\sigma~\sqrt{3}}{V_{\rm{A}}},\qquad \rm{and} \qquad
m_{\rm{s}}=\frac{\sigma~\sqrt{3}}{\textit{c}_{\rm{s}}}.
\label{m}
\end{equation}
Here $V_{\rm{A}}$ is the Alfv\'{e}n velocity and $c_{\rm{s}}$ is the sound velocity, and $\sigma$, the turbulence velocity, can be estimated using
 $\sigma=\Delta V_{\rm{turb}}/\sqrt{8~ln2}$.
By considering a mass $\mu_{\rm{H_{2}}}=3.8\times10^{-24}~\rm{g}$ and the relation between the velocity and the temperature
of a gas, we can write $V_{\rm{A}}$ and $c_{\rm{s}}$ in terms of 
$|B|$, $n_{\rm{H_{2}}}$, and the kinetic temperature of the gas $T_{\rm{k}}$
\begin{equation}
\frac{V_{\rm{A}}}{[\rm{km~s^{-1}}]}= 1542~\left(\frac{|B|}{[\rm{mG}]}\right)~\left(\frac{n_{\rm{H_{2}}}}{[\rm{cm^{-3}}]}\right)^{-\frac{1}{2}},
\end{equation}
\begin{equation}
 \frac{c_{\rm{s}}}{[\rm{km~s^{-1}}]}=0.0603~\left(\frac{T_{\rm{k}}}{[\rm{K}]}\right)^{-\frac{1}{2}}.
\end{equation}
So we get
\begin{equation}
\beta=3.058\cdot10^{-9}~\left(\frac{|B|}{[\rm{mG}]}\right)^{-2}~
\left(\frac{n_{\rm{H_{2}}}}{[\rm{cm^{-3}}]}\right)~\left(\frac{T_{\rm{k}}}{[\rm{K}]}\right),
\label{beta2}
\end{equation}
which, considering Eq.~\ref{zem} and $|B|=|B_{||}|/cos ~\langle\theta\rangle$, can thus be written as
\begin{equation}
\beta=3.058\cdot10^{-9}~\alpha_Z^{2} ~cos~\langle\theta\rangle ~\left(\frac{|\Delta V_{\rm{Z}}|}{[\rm{ms^{-1}}]}\right)^{-2}~
\left(\frac{n_{\rm{H_{2}}}}{[\rm{cm^{-3}}]}\right)~\left(\frac{T_{\rm{k}}}{[\rm{K}]}\right).
\label{beta3}
\end{equation}
By assuming $n_{\rm{H_{2}}}=10^{9}~\rm{cm^{-3}}$, $T\sim200$~K, which are the typical values in the \meth ~masing region, the 
weighted values for the unsaturated masers of the $\theta$ angle $\langle\theta\rangle=$78\d, and 
Zeeman-splittings $|\Delta V_{\rm{Z}}|=2.7$~\ms, we obtain
\begin{equation}
\beta=17.4 \cdot \alpha_Z^{2}.
\label{beta4}
\end{equation}
Although the value of $\alpha_Z$ is still uncertain, we can expect that this must be neither higher than the old value 
($\alpha_Z=0.049$~\kmsg, Vlemmings \cite{vle08}) nor lower than the new value ($\alpha_Z=0.005$~\kmsg, Vlemmings et al. \cite{vle11}). 
Therefore, it is reasonable that $\beta$ is between $4\cdot10^{-4}$ and $4\cdot10^{-2}$. Consequently the magnetic field is dynamically important 
in this massive star-forming region.
\begin{figure*}[ht!]
\centering
\includegraphics[width = 10 cm]{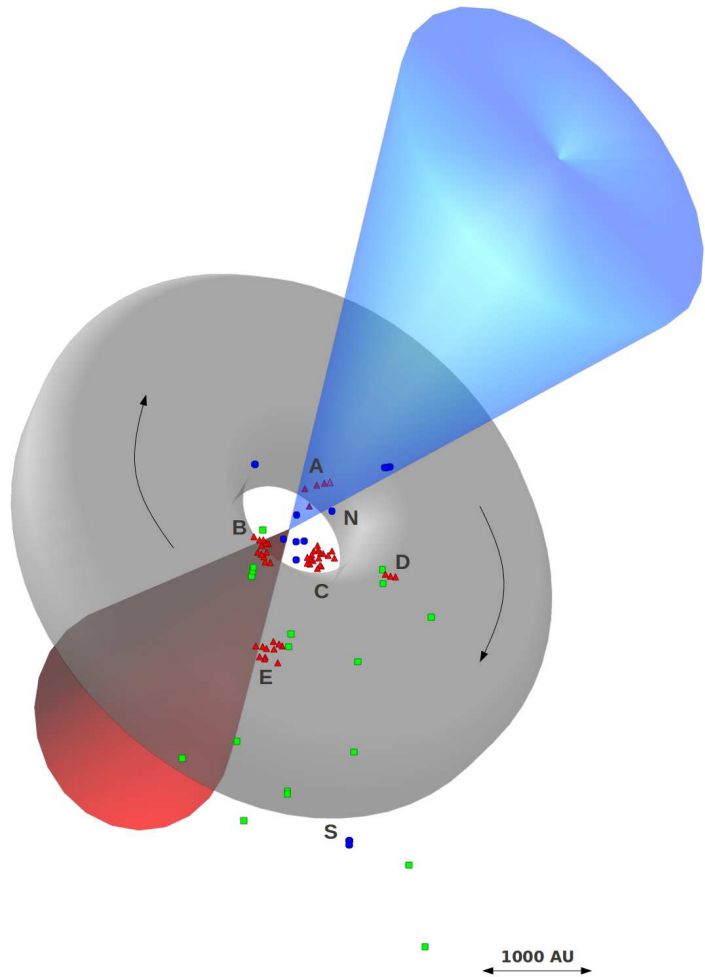}\\
\caption{The three-dimensional sketch of the massive star-forming region NGC7538-IRS\,1 as described in Sect.~\ref{sub3d}. The two cones are 
the red- and blue-shifted part of the large-scale molecular bipolar outflow (Kameya et al. \cite{kam89}), the donut is the torus 
suggested by Klaassen et al. (\cite{kla09}) perpendicular to the outflow and with i=32\d ~(De~Buizer \& Minier \cite{bui05}), the 
triangles, the circles, and the boxes are the methanol, \hdueo ~masers (present paper), and the OH masers (Hutawarakorn and Cohen \cite{hut03}),
 respectively. The outflow and the torus are transparent to allow the reader to see behind them.}
\label{3d}
\end{figure*}
\subsection{Structure of NGC7538-IRS\,1}
\label{sub3d}
\begin{figure}[th!]
\centering
\includegraphics[width = 9 cm]{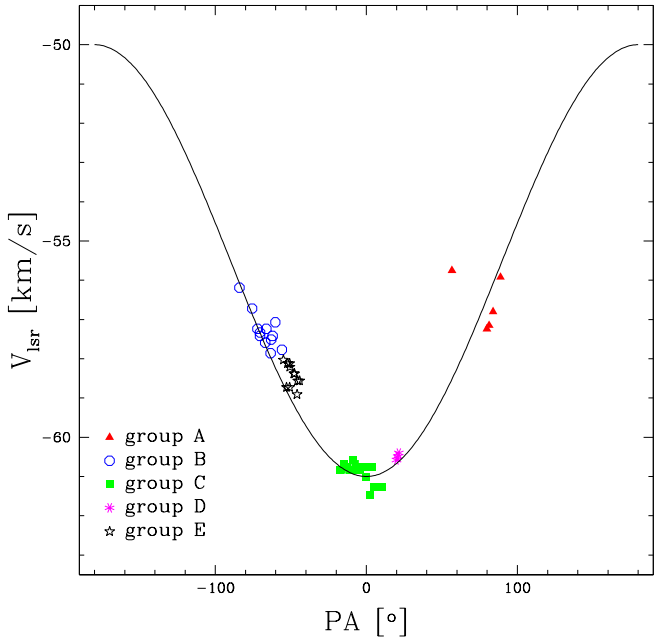}\\
\caption{Measured velocities of the \meth ~masers as a function of the position angle (PA) along the torus. The solid 
line shows the velocities of the torus reported by Klaassen et al. (\cite{kla09}) as function of their PA along the torus determined as
described in Sect.~\ref{sub3d}.}
\label{veldia}
\end{figure}

Recently, large-scale elliptical configurations have been detected in significant \meth ~masers around high-mass protostars. These masers 
are thought to trace molecular rings (e.g., Bartkiewicz et al. \cite{bar09}). However, most of these rings do not show signs of rotation, 
but instead the radial motions dominate, indicating that the masers are instead tracing infalling gas in the interface between
the torus and the flow, e.g. Cepheus~A~HW2 
(Torstensson et al. \cite{tor11}; Vlemmings et al. \cite{vle10}). In this light it is legitimate to reconsider whether the masers in 
NGC7538-IRS\,1 could trace similar gas.\\
\indent In this work we have observed velocity gradients in three \meth ~masers, i.e., E26, E33, and E40. Whereas the velocity gradient of
E26 has already been observed by other authors (e.g., Minier et al. \cite{min98}; Pestalozzi et al. \cite{pes04}), those of masers 
E33 and E40 have not been reported so far (Fig.~\ref{stream}). All velocity gradients of E26, E33, and E40 are equal to
0.02~\kms~AU$^{-1}$. Considering the orientation of the linear distribution of the maser spots of E26, E33, and E40, we find that all of them
seem to point toward a common center. This suggests that E26, E33, and E40 are likely tracing a gas with a radial motion, probably consistent
 with the interface of the torus and infalling gas. Consequently the maser spots of E26 are likely not tracing a Keplerian disk as proposed by
 Pestalozzi et al. (\cite{pes04}). The \meth ~masers of groups A, B, C, and D show a cone-shape distribution that opens to the north-west. This
 seems to indicate that we are looking at a regular structure, for instance the inner part of a torus. Klaassen et al. (\cite{kla09}) detected 
an almost face-on 5300~AU-torus
 perpendicular to the outflow with a clockwise rotation motion. If this torus is on the same plane of the disk supposed by 
De~Buizer \& Minier (\cite{bui05}), then its inclination angle must be i=32\d. By keeping in mind these new 
results, we can suppose that the \meth ~masers of groups A, B, and C are tracing the surface of the inner torus, possibly where 
the infall reaches the rotating structure. 
The masers of groups D and E trace the gas slightly farther out on the torus. In particular, group A is 
located to the opposite side of the torus w.r.t. the other three maser groups. 
In Fig.~\ref{3d} a three-dimensional sketch of the region as described above is shown.
The outflow sketched here is a simplified representation of the chaotic and complex multi-outflows structure observed toward NGC7538 and 
in particular around IRS-1 (e.g., Qiu et al. \cite{qiu11}; Klaassen et al. \cite{kla11}).\\
\indent In this scenario the \hdueo ~masers of group N are located parallel to the edges of the outflow, which show a blue-shifted
part northwestern and a red-shifted part southeastern (e.g., Kameya et al. \cite{kam89}; Hutawarakorn \& Cohen \cite{hut03}). 
In this case the \hdueo ~masers can be pumped by a shock caused by the interaction of the outflow with the infalling gas. The most
southern \hdueo ~masers, W06 and W07, might be associated either with the red-shifted part of the outflow or most likely
with another source, as also
suggested by the different orientation of the magnetic field in that region. Moreover, the OH masers located southward (Hutawarakorn \&
Cohen \cite{hut03}) might be pumped by the red-shifted part of the high-velocity outflow and those located westward by the blue-shifted 
part.\\
\indent This scenario is further supported by the similarity between the torus and outflow velocities and that of the three maser species.
In Fig.~\ref{veldia} we plot the \meth ~maser velocities as a function of the position angle (PA) along the torus. To determine the
PA of each individual \meth ~maser, we consider that every maser lies on an own circle that has the same orientation and inclination angle of the torus
 with the center located at the origin of the sketched outflow in Fig~\ref{3d}. Moreover, we plot in Fig.~\ref{veldia} an empirical function 
(solid line) based on the orientation and on the velocity field of the torus observed by Klaassen et al. (\cite{kla09}), i.e. the PA and velocity
of all the points of a circle that has the same size, orientation, and inclination of the torus. 
The maser velocities perfectly match the non-Keplerian velocity profile of the torus that is rotating clockwise (Klaassen et al. 
\cite{kla09}). This indicates that the \meth ~masers are likely related to the torus structure. 
Groups A and E appear to be systematically blue- and red-shifted w.r.t the torus velocities respectively, supporting the suggestion of infall 
motion (on the order of 1~\kms) along the line of sight.\\
\indent The velocities of the blue- and red-shifted part of the outflow are $V_{\rm{outf}}^{\rm{blue}}=-76$~\kms ~and 
$V_{\rm{outf}}^{\rm{red}}=-37$~\kms ~(Kameya et al. \cite{kam89}). The mean velocities of the \hdueo ~masers of group N and S
are $V_{\rm{H_2O}}^{\rm{N}}=-59.4$~\kms, excluding the masers W16 and W17, and $V_{\rm{H_2O}}^{\rm{S}}=-68.7$~\kms. While 
$V_{\rm{H_2O}}^{\rm{S}}$ again suggests that these masers might be associated with another source, the difference between 
$V_{\rm{H_2O}}^{\rm{N}}$ and $V_{\rm{outf}}^{\rm{blue}}$ indicates that the \hdueo ~masers are  
tracing the gas surrounding and being entrained by the outflow. The 1.6 and 1.7-GHz OH masers show 
velocity ranges $-61$~\kms$<V_{\rm{OH}}<-51$~\kms ~that agree well with $V_{\rm{torus}}$ and $V_{\rm{CH_3OH}}$, which might indicate an 
association with the torus rather than with the CO-outflow. The presence of OH and \meth ~masers in the same environments, 
also suggested by the linear polarization vectors of the two maser species (see Sect.~\ref{mfo}), is in contrast with the excitation 
model of Cragg et al. (\cite{grag02}), which predicts the inhibition of the 6.7-GHz \meth ~masers when 1.6-GHz OH masers arise.\\
\indent The orientation of the magnetic field derived by the linear polarized emission of the \meth ~masers at high 
angular resolution matches the scenario we presented above. Most likely the magnetic field is perpendicular to the linear polarization 
vectors for most of
the \meth ~maser features, but for group E the magnetic field is instead parallel to the linear polarization vectors (see Sect.~\ref{mfs}).
This is also suggested by considering the linear polarization vectors of the OH masers (see Sect.~\ref{mfo}). In this case the magnetic
 field is situated on the two surfaces of the torus, with a counterclockwise direction on the top surface. This interpretation is consistent 
with the signs of the magnetic field strengths determined from the Zeeman-splitting measurements of the \meth ~masers and OH masers, which show 
negative strength ($B=-2.0$~mG, Hutawarakorn \& Cohen \cite{hut03}) toward the region were groups B and E arises and positive toward groups
 A and D.\\
\indent Even if the \meth ~masers are not tracing the Keplerian-disk reported by Pestalozzi et al. (\cite{pes04}), the presence of a smaller 
disk is not ruled out. In the scenario proposed here, it would be located within the torus. Accordingly, we argue that a possible answer to the question 
about the morphology of NGC7538-IRS\,1 is that the \meth ~masers are tracing a gas close to the torus that is falling toward the connection 
region between the torus and the disk as was observed in other protostars (e.g., Vlemmings et al. \cite{vle10}).\\
\indent The new scenario described in this paper is determined by considering the entire \meth ~maser region 
and not just the masers of group A. Consequently, this scenario is different both from ``Scenario A'' and from ``Scenario B''
suggested by Kraus et al. (\cite{kra06}), who considered the results obtained by De Buizer \& Minier (\cite{bui05}) and by Pestalozzi et al. 
(\cite{pes04}) for the \meth ~masers of group A, respectively. In ``Scenario A'' the \meth ~masers of group A might resemble either clumps in 
the cavity of the outflow or recent ejecta from the outflow, while in ``Scenario B'' group A trace a disk and the authors proposed a jet precession
 model to explain the asymmetry of the NIR emission. However, in our scenario the asymmetry of the NIR emission might reflect the innermost
walls of an outflow cavity as explained in ``Scenario A'', but with the masers that are tracing an infalling gas rather than the outflow cavity or 
a recent ejecta from the outflow.\\
\indent Two questions remain: what is the radio continuum emission shown in Fig.~\ref{pos}? How can this scenario 
be consistent with the changing of the position angle of the outflow reported in Campbell (\cite{cam84})? De~Buizer \& Minier 
(\cite{bui05}) argue that the radio continuum emission is consistent with radio emission arising from a photoevaporated disk wind and Kameya et al.
 (\cite{kam89}) suggested that the rotation of the outflow stems from a disk precession.
\section{Conclusions}
The massive star-forming region NGC7538-IRS\,1 has been observed at 22-GHz in full polarization spectral mode with the VLBA
and at 6.7-GHz with the EVN and MERLIN to detect linear and circular polarization emission from \hdueo ~and \meth ~masers,
 respectively. We detected 17 \hdueo ~masers and 49 \meth ~masers at high angular resolution. We have measured
Zeeman-splitting for three \meth ~masers ranging from -2.7~\ms ~to +2.7~\ms. No significant magnetic field strength has been measured
from the \hdueo ~masers. Furthermore, we have also shown that the masers of NGC7538-IRS\,1 are all consistent with a torus-outflow 
scenario. Here the \meth ~masers are tracing the interface between the infall and the large-scale torus, 
and the \hdueo ~masers are related to the blue-shifted part of the outflow. The \hdueo ~masers of the southern group are thought to be associated
with another source. The magnetic field is situated on the two surfaces of the torus with a counterclockwise direction on the top
surface.\\

\noindent\small{\textit{Acknowledgments.}
We wish to thank an anonymous referee for making useful suggestions that have improved the paper.
GS, WHTV, and RMT acknowledge support by the Deutsche Forschungsgemeinschaft (DFG) through the Emmy 
Noether Reseach grant VL 61/3-1. G.S. and WHTV thank Pamela D. Klaassen for the very useful discussion. G.S. thanks Ramiro 
Franco-Hern\'{a}ndez for kindly providing the VLA continuum image.}

\end{document}